\documentclass[a4paper,11pt]{article}
\pdfoutput=1

\usepackage{jcappub}
\usepackage{caption}
\usepackage{subcaption}
\usepackage{gensymb}
\usepackage{amsmath}
\usepackage{amsfonts}
\usepackage{amssymb}
\usepackage{mathtools}
\usepackage{graphics}
\usepackage{citesort}
\usepackage{graphicx}

\newcommand{\myeqref}[1]{Eqn.~\eqref{#1}}
\newcommand{\mysecref}[1]{Sec.~\ref{#1}}
\newcommand{\myrefref}[1]{Ref.~\cite{#1}}
\newcommand{\intensityunit}{\ensuremath{\mbox{cm}^{-2}~\mbox{s}^{-1}~\mbox{sr}^{-1}~\mbox{GeV}^{-1}}}

\newcommand{\convint}{\mathop{\mathrlap{\,\bigstar}\int}}

\title{One-point fluctuation analysis of the high-energy neutrino sky}
\author[a]{Michael R. Feyereisen,}
\author[b,a]{Irene Tamborra}
\author[a]{and Shin'ichiro Ando}

\affiliation[a]{GRAPPA Institute, University of Amsterdam, Science Park
904, 1098 XH Amsterdam, Netherlands}
\affiliation[b]{Niels Bohr International Academy, Niels Bohr Institute,
Blegdamsvej 17, 2100 Copenhagen, Denmark}

\emailAdd{m.r.feyereisen@uva.nl}
\emailAdd{tamborra@nbi.ku.dk}
\emailAdd{s.ando@uva.nl}

\keywords{Neutrino astronomy, ultra high-energy photons and neutrinos, gamma-ray theory, particle acceleration.}

\abstract{We perform the first one-point fluctuation analysis of the
high-energy neutrino sky. This method reveals itself to be especially
suited to contemporary neutrino data, as it allows to study the
properties of the astrophysical components of the high-energy flux
detected by the IceCube telescope, even with low statistics and in the
absence of point source detection. Besides the veto-passing atmospheric
foregrounds, we adopt a simple model of the high-energy neutrino
background by assuming two main extra-galactic components: star-forming
galaxies and blazars. By leveraging multi-wavelength data from
{\it{Herschel}} and {\it{Fermi}}, we predict the spectral and
anisotropic probability distributions for their expected neutrino counts
in IceCube. We find that star-forming galaxies are likely to remain a
diffuse background due to the poor angular resolution of IceCube, and we
determine an upper limit on the number of shower events that can
reasonably be associated to blazars. We also find that upper limits on
the contribution of blazars to the measured flux are unfavourably
affected by the skewness of the blazar flux distribution. One-point
event clustering and likelihood analyses of the IceCube HESE data
suggest that this method has the potential to dramatically
improve over more conventional model-based analyses, especially for the
next generation of neutrino telescopes.}

\begin{document}

\maketitle

\newpage

\section{Introduction}

In 2013, the IceCube Collaboration reported an excess of high-energy
neutrinos over the atmospheric neutrino
background~\cite{Aartsen:2013bka,IC:2yr,IC:3yr,IC:4yr,Aartsen:2016oji}.
The spatial distribution of these events, consistent with isotropy and with no significant clustering, may suggest an extragalactic origin of the detected neutrinos~\cite{Anchordoqui:2013dnh,Aartsen:2014ivk,Neronov:2015,Leuermann:2016oxu}.
Besides a Galactic
contribution~\cite{Gaggero:2015,AhlersMuraseGalactic}, various
extragalactic astrophysical sources have been suggested as factories of
the IceCube neutrinos, e.g.~star-forming galaxies
(SFG)~\cite{loeb2006cumulative,He:2013cqa,Liu:2013wia,muraseahlerslacki2013,Tamborra:2014xia,Anchordoqui:2014,Chakraborty:2015sta,Senno:2015tra},
active-galactic
nuclei~\cite{PKS1424418,Kalashev:2013vba,Stecker:2013fxa,Murase:2014foa,Dermer:2014vaa,Padovani:2015mba,Wang:2015woa,Hooper:2016,Padovani:2016wwn},
galaxy clusters~\cite{Murase:2012rd,Zandanel:2014pva}, sources dim or
scarcely visible in
photons~\cite{Murase:2013ffa,Liu:2012pf,Tamborra:2015qza,Tamborra:2015fzv,Senno:2015tsn,Fujita:2015xva,Kimura:2014jba}
as well as more exotic dark matter
decays~\cite{Feldstein:2013kka,Esmaili:2013gha,Zavala:2014dla,Murase:2015gea}.
Recent work employing accurate statistical analysis as well as
up-to-date gamma-ray data-sets places strong constraints on some of
the proposed sources~\cite{Ando:2015bva,MuraseWaxman,BechtolSFG,Murase:2015xka}.
In this study we are interested in the joint contribution of multiple source populations to the observed extragalactic neutrino flux.

Given the paucity of the high-energy neutrino data, it is important to
extract as much information as we can from them. We here aim at
exploiting the full probability distribution of the currently available
neutrino data-set by employing a one-point fluctuation analysis
\cite{DMpaper,MalyshevHogg2011,lee2009Microhalo,Lee:2015fea,Zechlin:2015wdz,Lisanti:2016jub}. We
first model the high-energy neutrino sky in a simple data-driven way, by
assuming that neutrinos from SFGs and from blazars constitute the main
bulk of the observed IceCube flux, other than the atmospheric
background. The IceCube HESE data are then compared directly to our
model predictions. Our one-point analyses show that the specific model
of SFGs and blazars, carefully extrapolated from {\it Herschel} and {\it
Fermi} data, is insufficient to explain the IceCube astrophysical
excess. Our likelihood analysis suggests that the discrepancy can be
explained by missing un-modelled components that are likely of
astrophysical origin.

In addition to our analysis of the HESE data, the probability
distribution of the individual neutrino counts allows us to make
detection forecasts of these astrophysical populations as point sources
above diffuse backgrounds that they themselves generate. This extreme-value analysis suggests that a
detector with the IceCube angular resolution would not be likely to detect SFGs as point sources above the background of blazars and of other SFGs. On the other hand, blazars are sufficiently rare sources that they will not constitute a background to themselves. Instead, the skewness of the blazar flux distribution biases results derived from population averages by a non-negligible factor compared to the full distributional result, which we compute.

The paper is organised as follows. In \mysecref{sec:P1F}, we present
our data-driven modelling of the extragalactic and atmospheric neutrino flux.
In \mysecref{sec:PFC}, we predict what IceCube should observe on Earth as
a consequence of the adopted astrophysical
models and characterise the flux distributions of star-forming galaxies
and blazars, arguing that they are sufficiently skewed to bias results
on unresolved source contributions to the diffuse backgrounds. In \mysecref{sec:method} we present a few of the techniques available in one-point analyses, and in
\mysecref{sec:results}, we apply these techniques and expose the results of our analyses.
The systematics of this study are discussed in \mysecref{sec:caveats}, and
our findings are summarised in \mysecref{sec:concl}. Further materials
complementing the methodological discussions are reported in the Appendices.

\section{Distributional models of neutrino fluxes \label{sec:P1F}}

In this Section we describe the inputs we
used to model the neutrino emission from SFGs and blazars. We also derive
the flux probability distributions of single sources drawn randomly from these populations.

In this study we will consider the energy-differential particle fluxes $F$ (in units of $\mbox{cm}^{-2}~\mbox{s}^{-1}~\mbox{GeV}^{-1}$) of various sources. Specifically, we will be considering the statistics of the flux in
individual pixels, and to some extent we will be treating fluxes-per-pixel as equivalent to intensities $I=
F/\Omega_\mathrm{pix}$ (in units of $\intensityunit$).

In addition to
these energy-differential quantities, the gamma-ray studies we use to inform out models often work with
 fluxes $S_\gamma$ (in units of $\mathrm{cm}^{-2}~\mathrm{s}^{-1}$) integrated
over a certain energy range $[E_\mathrm{min},E_\mathrm{max}]$. Integrated neutrino fluxes $S_\nu$ will also be relevant in \mysecref{sec:PC/formalism}. For a differential flux with fixed spectral index $\Gamma$ (i.e., $F\propto
E^{-\Gamma}$), $S$ is related to $F$ by
\begin{equation}
S = F \times \frac{E_\mathrm{max}^{1-\Gamma}-E_\mathrm{min}^{1-\Gamma}}{(1-\Gamma)~E^{-\Gamma}}\ .
\label{eq:diffflux_conversion}
\end{equation}
Hence, when the spectral index over this energy range is known, $F$ and $S$ are also effectively interchangeable, and we can extrapolate GeV gamma-ray fluxes to their TeV--PeV
gamma-ray counterparts. From here we can further extrapolate their corresponding neutrino fluxes assuming $pp$ or $p\gamma$ interactions. Note that we will assume a single injection spectral index $\Gamma$ as
representative of the whole source population for simplicity.
In \mysecref{sec:caveats/marg}, we will discuss the systematics incurred by
employing such an approximation.
Notational preferences for $F$, $S$, or $I$ throughout the text are mainly to emphasise whether or not we are assuming a fixed pixel size ($I$), a fixed energy range ($S$), or neither ($F$).

\subsection{Star-forming galaxy fluxes from the \emph{Herschel} data \label{sec:P1F/SFG}}
We now introduce our model for the neutrino emission from star-forming galaxies. The probability distribution of their neutrino flux is also discussed.
\subsubsection{Flux model \label{sec:P1F/SFG/F}}

In a proton-rich astrophysical environment, the neutrino emission can be directly correlated to the gamma-ray emission \cite{AhlersHalzen,Anchordoqui:2013dnh}: 
\begin{equation}
\frac{1}{3}\sum_{\alpha=1}^6 E_\nu Q_{\nu,\alpha} = \frac{\kappa}{2} E_\gamma Q_\gamma\ , \label{eq:nugamma}
\end{equation} where $\alpha$ runs over (anti)neutrino flavours, $Q$ is
the energy-differential emission rate per source (in units of
$\mathrm{s}^{-1}~\mathrm{GeV}^{-1}$) and $\kappa= 2$ for hadro-nuclear interactions.
Using the direct relation between the neutrino and the gamma-ray
energies ($2 E_\nu=E_\gamma$) and integrating over source densities on
both sides of Eq.~(\ref{eq:nugamma}) to get the differential fluxes (in
units of $\mathrm{cm}^{-2}~\mathrm{s}^{-1}~\mathrm{GeV}^{-1}$), we have
$(1/6) \sum_{\alpha} F_{\nu,\alpha} = (\kappa/2) F_\gamma$. Since neutrino
oscillations push the flavour ratio towards 1:1:1 for extragalactic
sources, we can define the all-flavour neutrino and antineutrino flux as 
\begin{equation}
\label{eq:flux_conversion}
F_\nu \equiv \sum_{\alpha=1}^6 F_{\nu,\alpha} = 3 \kappa F_\gamma\ .
\end{equation}

Although we have a simple conversion between neutrino and gamma-ray
fluxes for hadronic sources, SFGs are barely resolved in gamma rays
(cf. e.g. \cite{FermiSFG2012,tangwang2014,pengwang2016,3FGL}).
Consequently, their neutrino flux distribution is derived following
\myrefref{Tamborra:2014xia}.
We adopt the {\it{Herschel}} infrared (IR) luminosity function,
$\Phi(L_\mathrm{IR},z)= d^2 N/(dV(z)\,d\log_{10}
L_{\mathrm{IR}})$~\cite{Herschel2013}, defined for the intrinsic
infrared luminosity $L_\mathrm{IR}$ and redshift $z$.
The IR luminosity function is connected to the gamma-ray luminosity
function $\Phi(L_\gamma,z)$ by an empirical correlation \cite{FermiSFG2012}
\begin{eqnarray}
\Phi_\gamma(L_\gamma,z) d \log L_\gamma &=& \Phi_\mathrm{IR}(L_\mathrm{IR},z) d \log L_\mathrm{IR}\ ,\\
L_\gamma(L_\mathrm{IR}) &=& 10^\beta \left(\frac{L_\mathrm{IR}}{10^{10}
      L_\odot}\right)^\alpha ~ \mathrm{erg~s}^{-1}\ ,
\end{eqnarray} 
where $\alpha =1.17\pm 6\%$ and $\beta = 39.28\pm 0.2\%$ and $L_\odot$ is the the solar luminosity. We will assume the best fit values of the above parameters in the following, though more rigorously we really should be marginalising over these uncertainties. The $0.2\%$ uncertainty on the normalisation exponent $\beta$ corresponds to an $18\%$ systematic uncertainty on the normalisation $10^\beta$. Meanwhile the uncertainty on the slope corresponds to a $\lesssim 2\%$ uncertainty on the normalisation for the values of $L_\mathrm{IR}$ at the edges of the domain of $\Phi(L_\mathrm{IR})$ \cite{Herschel2013}, and is correlated with $L_\mathrm{IR}$ itself. The combined systematic uncertainty on the extrapolation $L_\gamma(L_\mathrm{IR})$ for a single source (and so also on its neutrino flux $F_\nu = 3\kappa F_\gamma$) is then less than $\sim20\%$. For a further discussion of this systematic effect, see \mysecref{sec:caveats/marg}.

As discussed in Refs.~\cite{Herschel2013,Tamborra:2014xia}, the
luminosity function of IR galaxies can be decomposed into luminosity
functions for spiral (`normal') galaxies (NG), starburst galaxies (SB),
and star-forming galaxies hosting an obscured or low-luminosity AGN
(SF-AGN).
This last subpopulation is further divided into those having an energy
spectrum resembling the one of normal galaxies (SF-AGN (NG)) and those
more similar to starburst galaxies (SF-AGN (SB)); the redshift
evolutions of SF-AGNs is given in Table 2 of
\myrefref{Tamborra:2014xia}.
Moreover, SB-like galaxies usually have a harder spectrum than NGs
($\Gamma_\mathrm{SB} \simeq 2.2$ vs. $\Gamma_\mathrm{NG}\simeq 2.7$, see
\myrefref{Tamborra:2014xia,BechtolSFG} and references therein for more
details). In the following, we will only consider SB and
SF-AGN (SB) galaxies as main contributors to the high-energy neutrino
flux.

Since SF-AGNs represent the most abundant sub-class of SFGs, we also
computed the flux distribution of SF-AGN (NG) as a cross-check
($\Gamma_\mathrm{SF-AGN(NG)} = 2.7$). However, we find this subpopulation only
produces about 6\% of the SFG flux between 25~TeV and 5~PeV, well within systematic uncertainties, so this subpopulation has been neglected in what follows.

We assumed the energy-dependence of the $\gamma$-ray differential flux as an unbroken
power-law $\propto E^{-\Gamma_\mathrm{SB}}$ above 0.6~GeV \cite{Tamborra:2014xia}
and do not adopt an high-energy cutoff.
We will further discuss the effect of uncertainties on $\Gamma$ in
\mysecref{sec:caveats/marg}.

\subsubsection{Flux distribution \label{sec:P1F/SFG/P1F}}

For an SFG population composed of exactly $N=\int (dV/dz)
\Phi(L_\gamma,z) d\log L_\gamma dz$ sources, the luminosity function is
sufficient (under the assumption that these extragalactic sources are
isotropically distributed in a comoving cosmological volume element
$dV/dz$) to obtain the single source distribution:\footnote{Throughout
this paper, we denote probability distributions by $P(\cdots)$ and
distinguish them using the random variables that they describe, along
with subscripts if necessary. Conditional and parameterised
distributions are denoted as $P(\cdot|\cdot)$. The `exception' to this
convention is the Poisson distribution, denoted
$\mathcal{P}(\cdot|\cdot)$.}
\begin{equation}
\label{eq:P1Lz}
P_1(L_\gamma,z) = \frac{d^2N/dzdL_\gamma}{N} = \frac{dV}{dz}\frac{\Phi_{\gamma}(L_\gamma,z)}{N\ln(10)L_\gamma}\ .
\end{equation}
We use the Planck+WMAP cosmology in $dV/dz$ ($h=0.673$, $\Omega_\Lambda =0.685$,
$\Omega_\mathrm{m} =0.313$) \cite{Planck2013CosmoParams}.

For a population with a unique, fixed spectral index $\Gamma$ and
photons observed at energy $E_\gamma$ (i.e., emitted at various energies
$(1+z)E_\gamma$), the one-source gamma-ray differential flux
distribution is obtained by marginalising away the uncertainties on the
$(L_\gamma,z)$ of the source:
\begin{eqnarray}
P_1(F_\gamma | E_\gamma,\Gamma) = \iint dz dL ~ P_1(F_\gamma,L_\gamma,z|(1+z)E_\gamma,\Gamma) 
= \int dz ~ \left|  \frac{L_\mathrm{crit}}{F_\gamma} \right | P_1(L_\mathrm{crit},z)\ ,
\end{eqnarray}
where $L_\mathrm{crit}(F_\gamma,E_\gamma,\Gamma,z)$ is the $L_\gamma$ value obtained by the inversion of the differential flux model $F_\gamma(L_\gamma,\cdots)$ from \myrefref{Tamborra:2014xia} in which any attenuation during propagation is neglected. Inserting \myeqref{eq:P1Lz} then yields
\begin{equation}
P_1(F_\gamma | E_\gamma,\Gamma) = \frac{1}{|F_\gamma|} \int dz \frac{dV}{dz} \frac{\Phi_\gamma(L_\mathrm{crit},z)}{N \ln(10)}\ ,
\label{eq:LFP1F}
\end{equation} where uncertainties of $\Gamma_\mathrm{SB}=2.2$ are explicitly neglected. The effect of systematic uncertainties of $\Gamma$ on the mean flux in such a model has already been studied in \myrefref{Tamborra:2014xia}, and the systematic effects of statistical uncertainties on $\Gamma$ are discussed in \mysecref{sec:caveats/marg}. The SFG normalisation $N$ is effectively absorbed into the normalisation $\int dP =1$ of this single-source probability distribution, although it remains determined by the {\it Herschel} observations when we extrapolate this gamma-ray flux to neutrinos using \myeqref{eq:flux_conversion}.

\begin{figure}[h]
\centering
\includegraphics[scale=0.65]{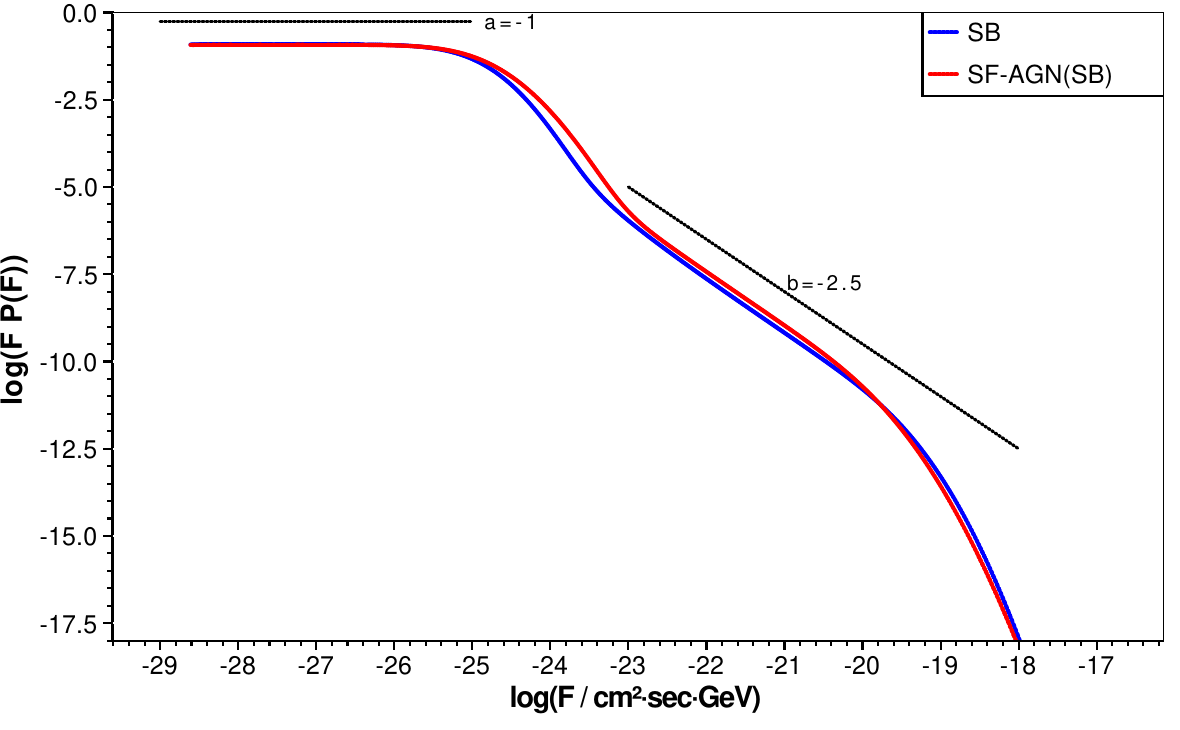}
\caption{Probability distribution $P_1(F_\nu)$ of the differential
 neutrino flux from a single star-forming galaxy at 100~TeV.
 The flux distributions of two SFG subpopulations are shown: SB (blue) and SF-AGN (red) \cite{Tamborra:2014xia}. Constant log-slopes corresponding to the limiting $1/|F|$ and the Euclidean behaviours ($a$ and $b$ respectively) are offset and quantified for convenience.}
\label{fig:starburstP1}
\end{figure}

For the high-flux tail, with contributions only from the nearby sources, the volume probed is very small and we expect an Euclidean scaling $F^{-2.5}$. The resulting $P_1(F_\nu)$ is then a broken power-law, up to corrections due to the redshift evolution of the SFG populations~\cite{Herschel2013}, as visible in Figure~\ref{fig:starburstP1}.

\subsection{Blazar fluxes from the \emph{Fermi} 2FHL catalogue \label{sec:P1F/2FHL}}

The second class of extragalactic neutrino sources in our model are blazars. In what follows we introduce our model to estimate their neutrino emission the basis of  based on their observed $\gamma$-rays spectra. The probability distribution of their flux is briefly introduced.

\subsubsection{Gamma-ray flux model \label{sec:P1F/2FHL/Fg}}

To construct our data-driven model of blazars, we rely on the source count distribution $dN/dS_\gamma$ of the 
Second Catalog of Hard \emph{Fermi}-LAT Sources (2FHL) \cite{2FHL,FermiResolvingAbove50GeV}. The 2FHL sources are mostly blazars, specifically BL Lacs.
One may justify this claim by extrapolating the observed contributions
from different blazar populations at high energy ($\sim$54\% BL Lac and
$\sim$16\% other blazar sub-populations \cite{2FHL}) to the unassociated
and unresolved sources. However, before extrapolating this gamma-ray flux into a neutrino flux, we must extrapolate it up to the IceCube energy range.

The 2FHL catalogue has substantially different properties from 2LAC, a 2FGL-based catalogue more commonly used in blazar-neutrino studies \cite{2LAC, Glusenkamp}. 2LAC is of a higher purity than 2FHL (97\% of sources are blazars of various subclassifications); however these sources are observed using gama-rays at energies $100~\mbox{MeV} - 100~\mbox{GeV}$, while the 2FHL is based on data between $50~\mbox{GeV} - 2~\mbox{TeV}$ \cite{2LAC, 2FHL}. Consequently, extrapolating the gamma-ray flux of 2LAC sources to their neutrino flux above 10 TeV is more dangerous than extrapolating the $\gamma$-flux of 2FHL sources.

This is relevant because the spectrum of these sources is very different in the two catalogues. BL-Lacs in the 2LAC have $\Gamma < 2.2$, but appear much softer in the higher energy range of 2FHL. For example, Fig.~10 of \myrefref{2FHL} shows the distribution of $\Gamma$ in a sample of BL Lacs shared between 2FHL and lower energy catalogues, with the clear trend that these sources' indices get softer at increasing energy, with median $\Gamma > 3$ in the 2FHL. This softening is observed despite the larger fraction of HSP (`hard') to LISP (`soft') blazars in 2FHL than in catalogues at lower energies \cite{2FHL}, which suggests the unresolved sources we want to model are even softer. This spectral behavior is  consistent with the observation that the spectral energy distributions (SEDs) of individual blazars are concave functions. The gamma-ray spectrum approaching PeV energies might be expected to be even softer than those of the 2FHL.

Despite this evidence that the blazar index is $\Gamma>3$ at higher energies, we nevertheless assume a non-concave SED at high energies, using \myeqref{eq:diffflux_conversion} with $\Gamma_\mathrm{2FHL}=2.5$. We make this simplifying assumption not only since we expect the flux from a population with uncertain $\Gamma$ to be dominated by the hardest sources (cf. \mysecref{sec:caveats/marg}) and for ease of comparison with existing studies (e.g. Refs.~\cite{Glusenkamp,aartsen:2014muf}), but also because this harder-than-expected extrapolation will result in an optimistic estimate of the contribution from blazars in the 2FHL (and hence in overconservative significances in our one-point fluctuation analyses in \mysecref{sec:results}). We will further discuss the effect of uncertainties on $\Gamma$ in \mysecref{sec:caveats/marg}.

\subsubsection{Neutrino flux model \label{sec:P1F/2FHL/Fnu}}

Now that we can extrapolate the gamma-ray flux between $50~\mbox{GeV} - 2~\mbox{TeV}$ to higher energies, we want to turn it into a neutrino flux. In this case, \myeqref{eq:nugamma} does not apply. We adopt instead the following relation from \myrefref{Padovani:2015mba,Dimitrakoudis:2013tpa,Petropoulou:2015upa} for the (all-flavors) neutrino flux:\begin{equation}
E^2_\nu F_\nu(E_\nu) = \left[\int_{10\,\mathrm{GeV}}^\infty E_\gamma F_\gamma dE_\gamma\right] \frac{Y}{0.9} \left(\frac{E_\nu}{E_{\nu,\mathrm{peak}}}\right)^{1-s} \exp\left(-\frac{E_\nu}{E_{\nu,\mathrm{peak}}}\right)
\label{eq:simplifiedblazars}
\end{equation} where $E_{\nu,\mathrm{peak}}\approx10~\mbox{PeV}$ for typical 2FHL sources ($z=0.4$, $\nu^S=10^{16}\mbox{Hz}$ \cite{2FHL}), and where $s=-0.35$ is used to obtain the denominator factor of $0.9$ in the normalisation \cite{Padovani:2015mba}. $Y$ is a parameter absorbing the details of the particle physics interactions in BL Lacs: the observed gamma-ray flux is mostly leptonic when $Y<1$, and mostly due to synchrotron emission from $p\pi$ when $Y\sim3$. The value $Y=0.8$ was chosen for ease of comparison with \myrefref{Padovani:2015mba}, though their discussions suggests smaller values of $Y$ may be more consistent with IceCube upper limits at the highest energies. This choice of a large $Y$ may therefore slightly overestimate the neutrino flux due to 2FHL sources, which will again result in overconservative significances for the discrepancies between our model and the HESE data we will dicsuss in \mysecref{sec:results}.

We can convert the integrated energy flux above 10 GeV in \myeqref{eq:simplifiedblazars} to the integrated particle flux in the 2FHL energy range, $S_\gamma$, using \myeqref{eq:diffflux_conversion}. The term in square brakets above becomes 
$\left[\int_{10\,\mathrm{GeV}}^\infty E_\gamma F_\gamma dE_\gamma\right] = S_\gamma (1-\Gamma)/(2-\Gamma) [-(10\;\mathrm{GeV})^{2-\Gamma}]/[(2\;\mathrm{TeV})^{1-\Gamma}-(50\;\mathrm{GeV})^{1-\Gamma}]$.
Thus we have \begin{equation}
F_\nu \propto S_\gamma E_\nu^{-(1+s)} \exp(E_\nu/E_{\nu,\mathrm{peak}})\ , \label{eq:linear_conversion}
\end{equation} with a predetermined proportionality constant that depends on the best-fit gamma-ray slope $\Gamma_\mathrm{2FHL}=2.5$ from \myrefref{FermiResolvingAbove50GeV}.
The log-derivative $\partial \ln F_\nu / \partial \ln E_\nu$ gives an energy-dependent neutrino spectrum $F_\nu\propto E_\nu^{-(1+s(E_\nu))}$ which softens as the energy increases, $s(E_\nu) = s + E_\nu/(10~\mbox{PeV})$. Note that the neutrino spectrum $s$ is different from the gamma-ray spectrum $\Gamma$ in our phenomenological model. A more accurate modeling of the microphysics may lead to more accurate predictions for $s$, but this goes beyond the demonstrative scope of our work.

\subsubsection{Flux distribution \label{sec:P1F/2FHL/P1F}}

These extrapolations $S_\gamma^\mathrm{2FHL} \to F_\nu$ are only the first step in determining the probability distribution $P_1(F_\nu|E_\nu)$ of the flux of any single source in the 2FHL. In terms of the number distribution $dN/dS_\gamma$ of sources resolved by {\it Fermi} in a flux range $[S_\gamma, S_\gamma+dS_\gamma]$, the single-source flux probability density is 
\begin{equation}
P_1(S_\gamma) = \frac{1}{N}\frac{dN}{dS_\gamma}\ .
\end{equation} A Monte-Carlo incorporating the \emph{Fermi} detection efficiency was used in
\myrefref{FermiResolvingAbove50GeV} to obtain the intrinsic
$dN/dS_\gamma$ of the 2FHL (i.e., the $dN/dS$ extrapolated below the detection threshold). In this extrapolation the nomalisation $N$ is implicitly determined by the faintest-source flux
cutoff $S_{\gamma,\mathrm{min}}=10^{-13}~\mbox{deg}^{-2} ~\mbox{cm}~\mbox{s}$, chosen to self-consitently reproduce the best-fit diffuse flux
observed by \emph{Fermi}~\cite{FermiResolvingAbove50GeV}. This flux
distribution, taking the form of a broken power-law, is a data-driven
model of these \emph{Fermi} sources, without any attempt at
discriminating subpopulations in the catalog and without consideration
of the physics which gives rise to these gamma rays. To compute $P_1(F_\nu|E_\nu)$ from $P_1(S_\gamma)$, notice that the flux conversion \myeqref{eq:linear_conversion} is effectively just a linear rescaling of the flux by a known term that depends on energies, on the spectral indices $\Gamma_\mathrm{2FHL}$, $s(E_\nu)$, and on the fixed quantities $Y,E_{\nu,\mathrm{peak}},s(10\,\mathrm{PeV})$.

\subsection{Atmospheric (cosmic ray) foregrounds \label{sec:atmospheric}}

\begin{figure}
\centering
\includegraphics[scale=0.65]{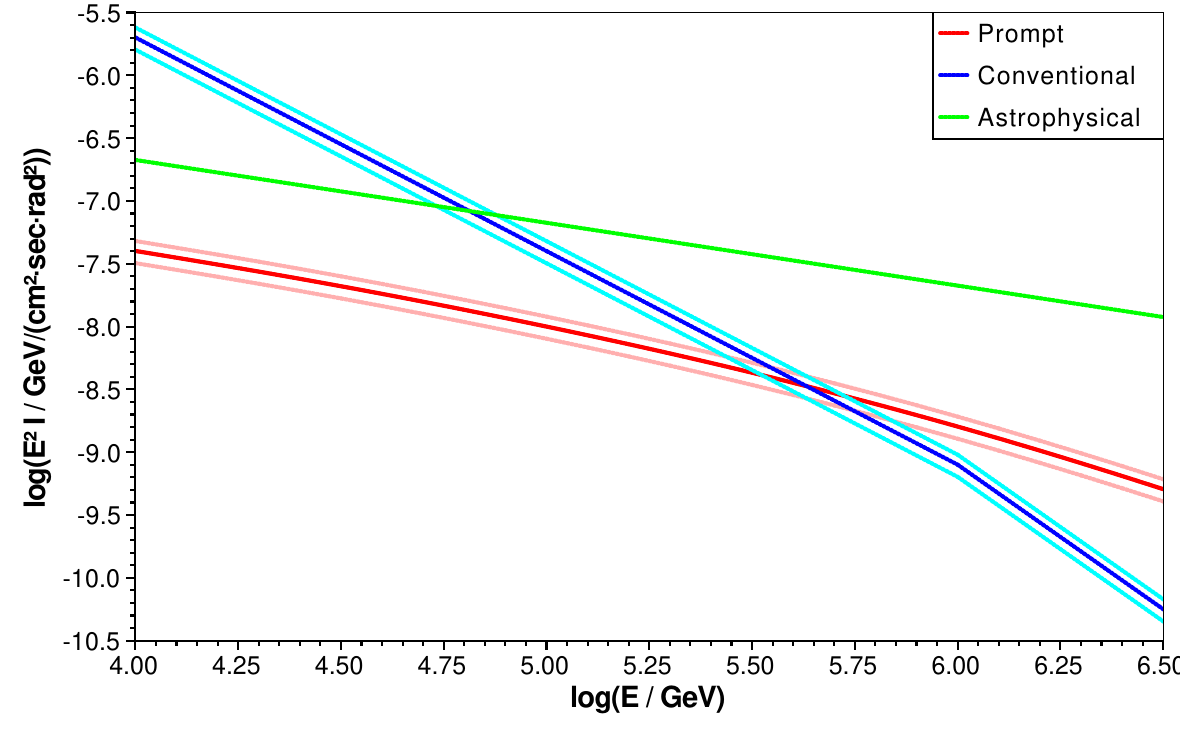}
\caption{Intensity $E_\nu^2 I_\nu(E_\nu)$ of the conventional (blue) and prompt (red) atmospheric contributions, as a function of energy. The $2\sigma$ bands shown here correspond to the intrinsic atmospheric variability $P(I_\nu|E_\nu)$ assumed in this analysis. These contributions are contrasted to the best-fit flux to the IceCube data from \myrefref{IC:astroflux} (green). In addition to these neutrino foregrounds, we also consider the veto-passing muon background (cf. main text).}
\label{fig:atmospheric}
\end{figure}

Before continuing our discussion of the flux distributions of extragalactic
astrophysical sources, we introduce the atmospheric foregrounds from
which these astrophysical contributions must be
extricated~\cite{IC:2yr}.
Atmospheric neutrinos produce an almost isotropic foreground with a soft
spectrum.
Our models of the conventional and prompt contributions are based on
\myrefref{Honda:2015} and \myrefref{Enberg:2008}, respectively.
We set the probability densities $P^\mathrm{atm}(I_\nu|E_\nu)$ of atmospheric all-flavour differential neutrino intensities $I_\nu= F_\nu/\Omega$ to Gaussians.\footnote{This can be justified, in the spirit of \mysecref{sec:PF/formalism}, by noting that this flux is the result of a very large number of cosmic ray interactions in the atmosphere, such that the central limit theorem may safely be assumed to hold for $P^\mathrm{atm}(I_\nu|E_\nu)$.}
The finesse of these distributions is chosen as
$\mu/\sigma = 10$ (i.e., a 10\% intrinsic variability in the atmospheric
intensity), the $2\sigma$ contours of which are respresented as the vertical width around the mean differential intensities in Figure~\ref{fig:atmospheric}.

The means of these distributions are determined as
follows. For the conventional contribution, the mean intensity is parameterised as 
\begin{equation}
\langle I_\nu (E) \rangle = 2 \times 10^{-14} \left(\frac{E_\nu}{10~\mbox{TeV}}\right)^{-\Gamma_\nu} \intensityunit\ ,
\end{equation} 
where the normalisation is set by the $\nu_\mu$ flux at 10~TeV in
\myrefref{Honda:2015} and the extra factor of two accounts for the
roughly equal flux of muon anti-neutrinos.
For the sake of simplicity, we neglect the anisotropic contributions to
the atmospheric flux. For the conventional contribution due to
$\nu_\mu$, this is mainly a zenith dependence at the South Pole
(cf. Fig.~7 in \myrefref{Honda:2015}). The spectrum $\Gamma_\nu$ is softer than the cosmic ray primaries by
$\Delta\Gamma=1$.
The cosmic ray knee is shifted down to about 1~PeV for neutrinos, such
that
\begin{equation}
\Gamma_\nu(E_\nu) = \begin{cases} 3.7 &\mbox{when~} E_\nu<1~\mbox{PeV} \\ 3.9 &\mbox{when~} E_\nu>1~\mbox{PeV} \end{cases}\  .
\end{equation}

In addition to this neutrino intensity, muon events passing the quality veto of the HESE data (cf. \mysecref{sec:pixsize} and \cite{IC:2yr}) were modelled by rescaling the conventional
flux by a factor of $4/3$, in accordance with the benchmark event rates
from the two-year study which claimed $\sim$4.5 and 6 events in
$\nu_\mu$ and $\mu^\pm$ respectively~\cite{IC:2yr}.

As for the prompt atmospheric contribution, we interpolate the average
$\nu_\mu+\bar{\nu}_{\mu}$ flux from \myrefref{Enberg:2008} as a function
of the energy, and add a rescaling factor of two to account for the
roughly equal muon and electron (anti)neutrino fluxes.
The enhanced prompt contribution from the proton intrinsic charm
\cite{HalzenWille,Laha:2016} is neglected given the opposite shifts in
flux from other updated QCD predictions (cf. e.g. \cite{2016prompt}),
and given the upper limits set in
Refs.~\cite{aartsen:2014muf,Aartsen:2016oji}.

The count distribution is then obtained by marginalising the flux
distribution into the detector response, with each pixel, energy bin,
and event topology treated independently, as will be described in the
next section.
Convolving all of these independent distributions gives the predicted
distribution of the total number of detected atmospheric neutrinos (and
veto-passing muons).
The average number of atmospheric counts between 25~TeV and 5~PeV in
this model is 27.9.
This can be further decomposed into 5.3, 9.7 and 12.9 events from prompt
neutrinos, conventional neutrinos, and veto-passing muons, respectively,
in rough agreement with an extrapolation of the two-year benchmark rates
from \myrefref{IC:2yr} to a four-year lifetime.
Poisson shot noise is the dominant source of uncertainty on these event
counts, but since we study the fluctuations themselves (statistically),
we are in principle sensitive to the assumed $P(I_\nu)$ rather than just
the mean $\langle I_\nu \rangle$.

\section{Flux and count distributions of single pixels \label{sec:PFC}}

In order to turn the above astrophysical models into predictions about
the data observed by IceCube, we must fold in some detector
characteristics (angular resolution, effective area, etc.), which will
be described in this section.
We also derive the total observed flux and count distributions.

\subsection{Size of a single pixel \label{sec:pixsize}}

In our analysis, we try to predict (from the data-driven models discussed above) both IceCube tracks and showers (cascades),
with pixel exposures constructed from a flavour, energy, and declination
dependent effective area tuned to the HESE
dataset~\cite{IC:2yr,IC:3yr,IC:4yr}.
This dataset consists of 54 events in the energy range [25~TeV, 5~PeV],
with interaction vertices contained within the detector: 39 showers, 14
tracks, and one coincident event not used in this study. Despite the stringent quality cuts, this neutrino dataset remains contaminated by veto-passing muons (cf. \mysecref{sec:atmospheric}), which contribute mostly but not exclusively to tracks (cf. Appendix~\ref{sec:flavaflav} and \myrefref{PalladinoFlavaflav}).

Since we are predicting probability distributions per pixel, we will make the simplifying assumption that pixel sizes are constant as
a function of the energy: roughly 30 degrees for showers and 1 degree
for tracks.
These correspond to rough estimates of the median angular resolution of
showers~\cite{IC:2yr} and contained tracks~\cite{Aartsen:2016oji} at the
energies considered in this study (25--5000~TeV).
These pixel sizes are used to bin the HESE events into 48 shower pixels
and 49152 track pixels, generated using HealPix~\cite{HealPix}; these  per-pixel counts will be directly compared to the predicted per-pixel count distributions (which include the flavour, energy, and declination-dependent HESE effective area) in \mysecref{sec:results}. Our study of probability distributions in pixels $\Delta \Omega$ rather than true one-{\it point} functions, although conceptually simpler, effectively ties us to this binned representation of the data. For a further discussion of binning and Healpix, see \mysecref{sec:caveats/method} and Appendix \mysecref{sec:declin}.

We emphasise that it is not in principle required to assume an energy-independent angular resolution to compute or study single-pixel fluctuation probabilities.
Furthermore, there is no methodological requirement to make pixels of
the same scale as the angular resolution. This choice is mostly for ease
of comparison with point source search studies in the
literature~\cite{IC:4yr,Aartsen:2016oji,MuraseWaxman,AhlersHalzen} and
our forecasts thereof in \mysecref{sec:results/rslvptsrc}.
Note that pixels must be at least as large as the angular resolution in
order to treat their fluxes as independent.

\subsection{Obtaining the total (multi-source) flux distribution \label{sec:PF}}

So far, we have been considering the flux distribution of a single source drawn at random from its population. However, in observations of abundant sources such as SFGs, there will be many sources in each of IceCube's pixels. The single-source quantity $P_1(F)$ therefore needs to be promoted to a single-pixel quantity $P(F)$. Here we discuss how to derive the latter from the former, and discuss some features of the blazar and SFG populations' per-pixel flux distributions.

\subsubsection{Formalism \label{sec:PF/formalism}}

The flux incident on a pixel is the sum of the fluxes of all the sources
in that pixel.
Given the distribution $P_1(F)$ of the spectral flux per source and the
distribution of the number of sources $N^\prime$ in a single given pixel, characterised by
the mean $\langle N^\prime \rangle = (\Omega_\mathrm{pix}/4\pi) N$, it
is straightforward to express the distribution $P(F)$ of the flux in a pixel in terms of the distribution $P(F|N^\prime)$ of the sum of the fluxes of $N^\prime$ sources: \begin{equation}
P(F) = \int P(F|N^\prime) P(N^\prime|\langle N^\prime \rangle) dN^\prime.
\end{equation} By independence of the sources and the algebra of random variables, the distribution of the
sum of their fluxes is given by the auto-convolution
\begin{equation}
P(F|N^\prime) = \underbrace{P_1(F) \star P_1(F) \star \cdots \star P_1(F)}_{N^\prime~\mathrm{times}}\ ,
\label{eq:autoconvolution}
\end{equation} 
where $f\star g$ denotes the convolution of the two probability distributions $f,g$.

The marginalised auto-convolution is as difficult to calculate as it is
straightforward to express, since the Fourier space techniques usually
applied to such compound Poisson distributions
\cite{scheuer1957statistical} develop numerical instabilities for
power-law-like $P_1(F)$ spanning many orders of magnitude.
We adopt instead the following Monte-Carlo strategy \cite{DMpaper}: 
\begin{itemize}
\item When the average number of sources per pixel $\langle N^\prime
      \rangle$ is too large to sample from $P_1(F_\nu)$ tails efficiently, we use the Central Limit Theorem to lump the faint sources into a diffuse
      Gaussian background and consider only the few brightest sources in
      that pixel.
\item When the average number of (bright) sources per pixel is small, we
      can repeatedly (i) realise this number $N^\prime$ from a Poisson
      distribution with the mean $\langle N^\prime \rangle$, (ii) draw
      that many samples from the single-source flux distribution, and
      (iii) add them all up. The histogram of the sums of fluxes
      approximates $P(F)$.
\end{itemize} 
In addition, the highest fluxes of $P(F)$ are known to converge to
$P_1(F)$ \cite{DMpaper}, which can be used to supplement the
Monte-Carlo-derived $P(F)$ with an analytical ``bright point-source''
tail.
Realising $N^\prime$ from a Poisson distribution without marginalising
over any uncertainties in $\langle N^\prime \rangle$ effectively
suppresses the variance of the statistically isotropic flux
(cf. \mysecref{sec:caveats/method}), to which one-point methods are sensitive as a signal rather than a background \cite{barcons1992confusion,Barcons15061994}.

\subsubsection{Discussion \label{sec:PF/discussion}}

\begin{figure}
\centering
\includegraphics[scale=0.75]{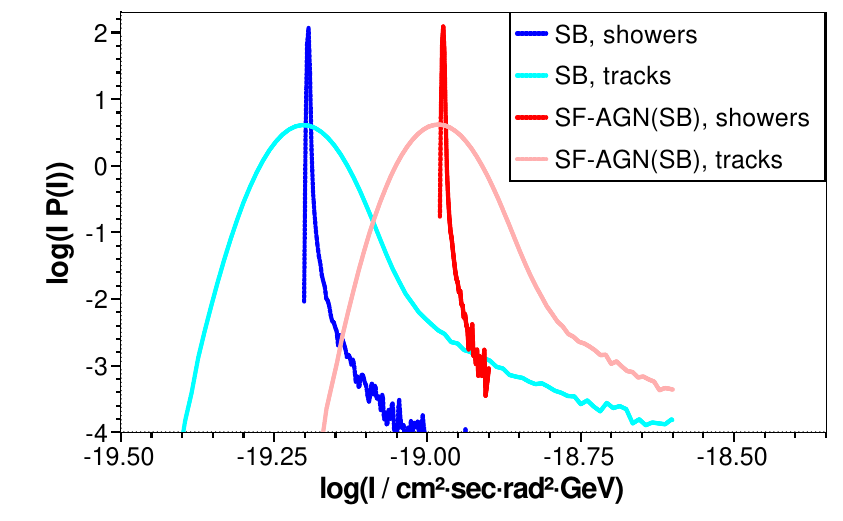}

\includegraphics[scale=0.75]{./2fhlpi.pdf}
\caption{{\it Top:} Probability distribution $P(I_\nu)$ of the SFG intensities as observable at 100~TeV. These distributions take the form of a Gaussian peak with a power-law tail. Starbursts are shown in blue (showers) and cyan (tracks), while SF-AGN (SB) are shown in red (showers) and pink (tracks). In each subpopulation, these peaks are much thinner in showers than in tracks as a consequence of the increased number of sources in larger pixels (cf. main text). {\it Bottom:} Probability distribution $P(I_\nu)$ of 2FHL source intensities at 100~TeV, in showers (dark green) and tracks (light green). These distributions are shown conditioned on there actually being a blazar in the pixel, so the absolute and relative normalisations are not visible in this figure. The cusp in tracks occurs at twice the minimum flux, it is the transition from one to two sources per pixel. Above this cusp, multiple sources contribute jointly to the flux, and a smooth bump begins to form.
}
\label{fig:variousPF}
\end{figure}

By applying \myeqref{eq:autoconvolution} to the SFG single-source flux distribution, we plot in the top panel of
Fig.~\ref{fig:variousPF} the probability distribution $P(I_\nu)$ of the SFGs
at 100~TeV for tracks and showers.
Note that the distributions have the form of a Gaussian with a power-law
tail.
These features correspond to the diffuse glow of a large number of
unresolved point sources, and to the few point sources with intensities
high enough to potentially be resolved individually~\cite{DMpaper}.
We postpone discussion of point-source-detectability prospects for these
populations until \mysecref{sec:results}. Here we focus on the
physical interpretation and consequences of the features of $P(F_\nu)$.

Although the single-source flux distribution is independent of the
normalisation of the SFG luminosity function, the total flux
distribution is sensitive to this normalisation via the average number
$\langle N^\prime \rangle$ of sources per pixel.
The uncertainties on the normalisations of the SB and SF-AGN (SB) luminosity
functions are not formally
considered as distributions to be marginalised away in this study. However, we discuss these uncertainties below and in \mysecref{sec:caveats/marg}.

The mean and variance of $P(F_\nu)$ are nothing other than a linear
rescaling of the mean and variance of $P_1(F_\nu)$ by a factor of $\langle
N^\prime \rangle$. For example, the relative locations of the SFG subpopulations in Fig.~\ref{fig:variousPF} are determined by the combination of two effects: firstly, there are roughly $(25\pm15)\%$ more members of the SF-AGN (SB) subpopulation than of the SB subpopulation in each pixel (according to the normalisations of the {\it Herschel} luminosity functions used in \mysecref{sec:P1F/SFG/F}); secondly, and more importantly, the mean flux of an individual SF-AGN (SB) is larger than the mean flux of an SB (consider Fig.~\ref{fig:starburstP1} between $10^{-25}$ and $10^{-20}$ {\ensuremath{\mbox{cm}^{-2}~\mbox{s}^{-1}~\mbox{GeV}^{-1}}). These two effects push the typical flux per pixel from SF-AGN (SB) sources slightly above that of SB sources in our model.

Moreover, we expect that the peak finesse increases with $\langle
N^\prime \rangle$, as a consequence of the $\sqrt{\langle N^\prime
\rangle}$ scaling of the finesse in the central limit theorem that gives
a Gaussian shape to the diffuse peak~\cite{DMpaper}.
This can be corroborated by looking, in Fig.~\ref{fig:variousPF} or in
Table~\ref{tab:truncMLE}, at the same populations in tracks and in
showers: tracks have a better angular resolution and therefore wider
diffuse peaks because $\langle N^\prime \rangle$ drops from
$\mathcal{O}(10^6)$ to $\mathcal{O}(10^3)$ in both SB and SF-AGN (SB).
A linear regression of the finesse of diffuse peaks from the four $P(F)$
distributions of Fig.~\ref{fig:variousPF} on their
respective $\sqrt{\langle N^\prime \rangle}$ yields a Pearson
$R^2=0.999$.

The locations of the peaks of these distributions are also slightly
offset among each other, the peak in showers is at slightly higher flux
than the peak in tracks (again, as visible in Fig.~\ref{fig:variousPF}
or in Table \ref{tab:truncMLE}). This is also a consequence of
convergence in the central limit theorem.
Indeed, the single-source distribution is power-law like and hence very
skewed, but the more sources we have in our pixel, the less we are
dominated by the individual source properties and the closer we get to
the population mean intensity---which is a quantity dependent on the
luminosity function, but independent on the angular resolution.
Since showers contain more unresolved sources than tracks, the diffuse
peak in showers is closer to the mean of these distributions than the
diffuse peak in tracks.

The flux distribution for 2FHL sources in showers (bottom panel of
Fig.~\ref{fig:variousPF}) is qualitatively similar to the flux
distributions of SFGs (top panel of the same figure), though its
power-law tail is much more prevalent; there are few enough sources per
pixel ($\langle N^\prime\rangle = 429$) that even the diffuse peak is
distinctly skewed.
This non-Gaussianity can, in principle, be exploited to characterise
diffuse backgrounds from unresolved sources, even though source number
density and source luminosity are degenerate at the level of
averages~\cite{MuraseWaxman}.
The small number of sources per pixel also means that the 2FHL peak is
much wider than the SFG peaks, by a factor exceeding an order of
magnitude (cf. Table~\ref{tab:truncMLE}).
This further corroborates the theoretically expected $\sqrt{\langle
N^\prime \rangle}$ scaling of the finesse.\footnote{Although this
scaling is a useful tool within a single population of unresolved
sources, across multiple source populations other population-specific
factors come into play and the explained variance $R^2$ decreases.}

The 2FHL distribution in tracks is informed by the tiny number of
sources per pixel ($\langle N^\prime\rangle=0.42$) and the sharp cutoff
imposed at lower fluxes in the single-source distribution in
\mysecref{sec:P1F/2FHL/P1F}.
The cusp in the bottom panel of Fig.~\ref{fig:variousPF}, located at
twice the minimum flux, corresponds to the discrete transition from one
to two sources per pixel, and below this cusp the distribution is
accordingly a pure power law (corresponding to a single source).
Above this cusp, multiple sources contribute jointly to the flux, and a
smooth bump begins to form; this bump becomes the diffuse peak when the pixel size increases.
The power-law tail sets in at higher flux when one of the blazars
dominates the flux of the others.
As in the case of SFGs, the most likely flux for tracks is still at
smaller flux than for showers.

\subsection{Obtaining the observed count distribution \label{sec:PC}}

The neutrino fluxes produce discrete event counts in our detector.
Having a detector model built into our pipeline means that we can
compare the distributional predictions of our astrophysical models
directly to the raw event count data in terms of count distributions $P(C)$.

\subsubsection{Formalism \label{sec:PC/formalism}}

At the bare minimum, a detector model consists of a pixel's
exposure and solid angle.
Having already accounted for the latter, the exposure can be constructed
by multiplying the IceCube livetime (roughly four years with a $95\%$
duty cycle) by its effective area $A$, which is flavour, energy, and
declination dependent~\cite{IC:2yr, IC:3yr,IC:4yr}.
We postpone discussion of the flavour and declination dependence to the
Appendices, and focus here on our distributional treatment of the energy dependence.

Using \myeqref{eq:diffflux_conversion} to convert from differential neutrino fluxes $F_\nu$ into integrated neutrino fluxes $S_\nu$, the distribution $P(S_\nu)$ (integrated over an energy bin $[E_{\nu,\mathrm{min}}, E_{\nu,\mathrm{max}}]$) can be made into a number of counts per pixel and per energy bin, by marginalising the flux distribution into the detector response, as \begin{equation}
P(C) = \int d\mu \mathcal{P}(C|\mu) P(\mu) ,\quad P(\mu) = \int \delta(\mu - S_\nu At) P(S_\nu) dS_\nu = \left. \frac{P(S_\nu)}{At} \right|_{S_\nu=\mu/At}.
\label{eq:pcpmu}
\end{equation} In this prescription, we first compute the distribution $P(\mu)$ of the mean number of counts, and assume these counts are the result of a Poisson process (completely uncorrelated) to obtain $P(C)$.

Assuming independence between multiple energy bins, we can merge bins by
convolving the distributions in each bin.
Indeed, the total integrated flux $S_\nu$ over a collection of bins is equal to the sum of the
integrated fluxes in each bin.
This extensive property of integrated fluxes/counts is useful to account
for the fact that the effective area $A$ is energy-dependent: we can
generate $P(C)$ in some large number of narrow energy sub-bins, where
the effective area varies across sub-bins but remains constant inside
each one, and then we can convolve the $P(C)$'s to merge the sub-bins
into a single bin.
We refer the interested reader to Appendix~\ref{sec:energy} for further
discussion of this construction.

As a tradeoff between wanting to exploit the spectrum and hoping to
circumvent the low statistics inherent in this endeavour, we generate
$P(C)$ in three final energy bins, with edges at
$[25,100,1000,5000]$~TeV.
In the real data, there are 34 events in the 20-100 TeV bin, with a
relative Poisson noise of $\sqrt{34}/34 \sim 17\%$ only marginally
larger than that of the full dataset ($\sqrt{53}/53 \sim 15\%$).
Of the remaining 19 events, only 3 events lie in the 1-5 PeV
bin~\cite{IC:4yr}.

Since we are working with relatively wide energy bins, we assume for
simplicity that the deposited energy and the neutrino energy are equal,
even though this is a poor approximation for tracks. We also neglect uncertainties due to the energy resolution (cf. \mysecref{sec:caveats/method}). These are
$\sim$5\% and $\sim$15\% systematic and statistical effects,
respectively~\cite{ICenergyreconstruction}.
Our treatment of the anisotropy of the exposure and its dependence on the incident flavour ratio are discussed in Appendices~\ref{sec:declin} and \ref{sec:flavaflav}.

\subsubsection{Discussion \label{sec:PC/discussion}}

The count distributions $P(C)$ for our extragalactic sources are, in
first approximation, Poisson distributions with means determined by the
``diffuse peak'' of $P(F)$, and the energy/declination-dependent
effective area in that pixel. Given the significant tail of $P(F)$, the
distribution has a skew, such that the location of the peak and the
location of the mean do not coincide. When we observe the sky, our
observation of event counts is biased by this skewness, as we are more
likely to observe the most probable number of counts than the mean
number. As discussed above, this bias is increasingly prominent as the
pixel size decreases or as the unresolved sources become rarer.

The skewness-induced reduction in the anticipated number of counts is automatically
accounted for by using the full $P(C)$ of \myeqref{eq:pcpmu} rather than the average $\langle
C \rangle$ of source populations. However, amongst other things, this weakens
upper limits determined from the population-average contributions of
these sources to the diffuse flux.
Such a weakening of upper limits derived using averages has already been
discussed in the context of dark matter constraints from the diffuse gamma-ray
background~\cite{DMpaper}.

Because SFGs are so abundant, this bias is at the percent level for these sources:
the average-derived limits on SFG contributions of, e.g., \myrefref{BechtolSFG} are only
a few percent weaker than the limits one would derive using the full distribution
--- but the fact that such studies of SFGs do not suffer from
this bias could not have been known without using their $P(F)$.

For 2FHL sources, on the other hand, the mean and mode of
$P(F)$ differ by factors of 0.4 in showers and 6.7 in
tracks, significantly reducing their anticipated count yield despite not
affecting their mean count yield. Knowledge of the total distribution $P(F)$ is, however, not necessary to
get a good approximation when sources are sufficiently rare that $P(F)\approx P_1(F)$.
For example, \myrefref{Glusenkamp} uses the blazar source count
distribution $ dN/dF \propto P_1(F)$ to derive its limits which (as a
consequence of $\langle N^\prime\rangle=0.42$) is a good approximation
to the full $P(F)$ in tracks: the $20\%$ upper limit on the blazar
contribution derived therein is not affected by this skewness.
Note however that the stacking procedure in \myrefref{Glusenkamp} increases the
effective $\langle N^\prime\rangle$ in the stacked pixel and thereby
deteriorates the quality of this approximation, see also our discussion
in \mysecref{sec:results/rslvptsrc/2FHL}.

Correcting for the skewness-induced bias just discussed using \myeqref{eq:pcpmu}, the average
number of counts $\langle C \rangle = \sum_{C} C P(C)$, cumulative over all energies and
declinations, in both tracks and showers, is then 2.2 events for SFGs
and 3.3 events for 2FHL sources. Notice, in the hard-spectrum blazar case, that this is approximately equal to the number of events for 1--5~PeV in the HESE data \cite{IC:4yr}. Even after subtracting the 28 atmospheric events predicted by our atmospheric model from the 53 actually observed, one expects roughly 20 of these events to remain unexplained by our fiducial model. Hence, the expected contributions from SFG and from 2FHL models are each about 10--15\% of the astrophysical flux, well below known upper limits \cite{BechtolSFG,Wang:2015woa,Glusenkamp}. The statistical significance with which we can say our data-driven model is incomplete (amongst other things we can learn from one-point functions) will be investigated in the next sections.

\section{Analysis (I): Methodology \label{sec:method}}

To show that our systematic conceptual approach is very general, we present in this study three different one-point analyses: a point source detection analysis, a probable clustering analysis, and a likelihood analysis based on the count distributions of individual pixels.

\subsection{Resolvability of point sources \label{sec:method/rslvptsrc}}

Let us consider the ideal limit of a telescope with fixed angular
resolution but infinite exposure.
The Poisson noise in such an instrument would be negligible, it would effectively
be sensitive to $P(F)$ directly rather than $P(C)$.
Even in this idealised situation, the finite angular size of a pixel
means that not all sources can be individually resolved:
the diffuse peak due to unresolved sources of a given population is
an intrinsic background to point sources of the same population. In what follows, we argue that even an ideal detector with the IceCube
angular resolution would be extremely unlikely to detect SFGs as point
sources.

A point source is basically just a localised flux observed in excess of a
predetermined threshold value $F^\mathrm{pt}$.
The probability that such a localised excess can be found in any single
pixel is given by the exceedance (complementary cumulative) distribution
of $P(F)$, and the typical number of excesses we expect to see in
$N_\mathrm{pix}$ pixels is
\begin{equation}
N_\mathrm{pt}(F^\mathrm{pt}) = N_\mathrm{pix} \int_{F^\mathrm{pt}}^{+\infty} P(F) dF \approx N_\mathrm{pix}  \frac{N(F>F^\mathrm{pt})}{N_\mathrm{MC}}\ ,
\label{eq:MCestimate}
\end{equation} 
where the latter has been obtained by estimating the exceedance
probability by Monte-Carlo sampling from $P(F)$.
We recall that $N_\mathrm{pix} = 48$ for showers and $N_\mathrm{pix}
\sim 5\times 10^4$ for tracks. The fluctuations around the expected number of sources are assumed to be Poissonian.

An analytic approximation to the exceedance probability, valid in the
high flux power-law tail where a single source dominates the flux in the
pixel, was derived in \myrefref{DMpaper}.
When this approximation matches the Monte-Carlo estimation above, we can
be relatively confident that the localised excesses correspond to single
astrophysical objects. However, there is a region between the diffuse peak and
the power-law tail where multiple bright sources jointly contribute to the
flux and might be confused for a single point source. Because of this possible confusion,
the number of localised excesses is always greater than the number of astrophysical point sources. The upper limits for the detection of
astrophysical sources we will determine using localised excesses are therefore conservative.
Also note that stacked searches \cite{Glusenkamp,MuraseWaxman} are intrinsically looking for
localised statistical excesses rather than individual point sources.

In order to study excesses above the diffuse background, we must characterise the Gaussian peak of $P(F)$.
To do this, we take the samples drawn from $P(F)$, and censor the values above the peak of the
distribution where non-Gaussianities due to the power-law nature of the
single-source distributions might arise.
We then fit the samples below the peak to a doubly truncated normal
distribution using the maximum likelihood estimators derived in
\myrefref{cohen1950}.\footnote{Our truncation points are (i) the flux
at which the distribution peaks, and (ii) a flux of $F=0$. We still
determine $\hat{\mu}$ from the Monte Carlo samples for the sake of
self-consistency, in case the truncation point (derived from an
interpolation of the samples) is not exactly at the distribution
peak. See \mysecref{sec:caveats} for a discussion of effects that
contribute to producing a non-Gaussian diffuse peak.}
The estimated mean $\hat{\mu}$ and standard deviation $\hat{\sigma}$ of the diffuse peak of each population
(reported in Table~\ref{tab:truncMLE}) can then be used to define flux
thresholds of localised excesses above the diffuse peak with various
signal-to-noise ratios, $F^\mathrm{pt}(\mathrm{SNR}) = \hat{\mu} + (\mathrm{SNR}) \hat{\sigma}$.

Using the complementary cumulative distribution of $P(F)$, these thresholds can be converted into the
exceedance probabilities associated to any given signal-to-noise ratio $\mathrm{SNR}$. These are larger than for a pure Gaussian because $P(F)$ is skewed. Indeed the rarer a given population of sources, the more skewed its $P(F)$ is (cf. \mysecref{sec:PF}) and therefore the more probable its exceedances are to have high signal-to-noise ratio.

In order to see such a source from a rare population, however, it must also be brighter than the diffuse backgrounds of all other source populations combined. Treating these $P(F)$ peaks as Gaussians comes with the benefit that the diffuse backgrounds due to multiple populations can easily be convolved into a single diffuse and isotropic extragalactic neutrino background with mean intensity $\sum_i\hat{\mu}_i$ and width $\sqrt{\sum_i (\hat{\sigma}_i)^2}$. Similar thresholds $F_\mathrm{pt}(\mathrm{SNR})$ may be defined for this total background, and the exceedances of individual populations above this total background may be forecasted (cf. \mysecref{sec:results/rslvptsrc}).

Heuristically, the skewness of $P(F)$ is due to barely-resolvable point sources in this infinite-exposure idealisation. Notice that decreasing the pixel size increases the skewness of $P(F)$ (cf. Fig.~\ref{fig:variousPF}), and so increases the typical signal-to-noise of excesses: barely-resolvable sources (e.g. the ``hot spots" of flux maps \cite{Bartels:2015aea,Aartsen:2016oji}) may become resolvable to future instruments. By extension, the excess skewness of $P(C)$, over the skewness of a Poisson distribution in IceCube, is related to the possible improvement in discovery potential of future point-source searches in an instrument with improved angular resolution, such as KM3NeT (ARCA)~\cite{KM3NeT} or IceCube-Gen2 \cite{ICGen2}.

\subsection{A ``Pointless" clustering analysis \label{sec:method/clustering}}

The IceCube collaboration has found no evidence for clustering by
looking for hot spots consistent with point
sources~\cite{IC:4yr,Aartsen:2016oji}.
But resolving point sources is not the only way we might see clusters of
events: in a detector with realistic exposure we can also exploit the
statistical properties of localised event clusters due to multiple bright but
unresolved sources or even shot noise fluctuations.

Given a fixed pixel size, the one-point function is the most
straightforward tool to study neutrino clustering.
Indeed, we can directly consider the ``average number of clustered
neutrinos per pixel'' or the ``rarity of a cluster of $N\ge2$ or more
events,''
\begin{equation}
\langle C \ge 2\rangle = \sum_{C=2}^{\infty} C 
 P(C),~~\mathfrak{C}(N) = \sum_{C=N}^\infty P(C)\ .
\end{equation}
In the ideal case that the data reproduce exactly a Poisson distribution
with a mean $\mu$, it is easy to show, e.g., that $\langle C \ge
2\rangle=\mu(1-e^{-\mu})$.
However, not only the na{\"i}ve analysis above would not account for the
different angular resolution of tracks and showers and the anisotropic
exposure, it would also eschew distributional information by using a
single $\mu$ value rather than the full $P(\mu)$ from
\myeqref{eq:pcpmu}.
One should expect two effects to emerge from the power-law tails of
astrophysical contributions: on the one hand, this tail increases the number
of clustered events; on the other hand, this tail contributes a skewness
that pushes the most likely values of the distribution to lower flux, as
part of a distribution with a fixed mean, resulting (after
marginalisation) in less event clustering overall~\cite{DMpaper}.

To go beyond the mean values, let us consider the following per-pixel $(p)$
clustering statistic with a model-dependence {\bf M} on the flux
distributions and the detector response from \mysecref{sec:PFC}:
\begin{equation}
\mathfrak{C}^{(p)} = \begin{cases}
\sum_{C=d^{(p)}}^\infty P(C|{\bf M}) &\mbox{if } d^{(p)} \ge 2 \\
1 & \mbox{otherwise}
\end{cases}~. \label{eq:clusteringC}
\end{equation} 
The total amount of clustering associated to a dataset $\{\forall p, d^{(p)}\}$ is then quantified by $\mathfrak{C} = \prod_p \mathfrak{C}^{(p)}$, and data sets with more clustering (given the same $P(C|{\bf M})$) will have a larger $-\ln \mathfrak{C}$. Since we care only about directional information in this test statistic, we need to treat coincident neutrinos of different energies as members of the same cluster. To do so, we follow the prescription of \mysecref{sec:PC/formalism} and Appendix~\ref{sec:energy} and convolve the count distributions of different energy bins to produce the $P(C)$ of \myeqref{eq:clusteringC}. The same logic applies to coincident tracks and showers: accurate track pixels were first coarse-grained into shower-sized pixels (cf. Appendix~\ref{sec:declin}), and then convolved with the shower pixel covering the same patch of sky.

A clustering analysis using $\mathfrak{C}$ is in a sense a
generalisation of the multiplet method applied to the IceCube data
in \myrefref{MuraseWaxman}.
In their analysis, the average number of sources producing $C\ge2$
tracks was computed from populations of ``effective standard candles,''
i.e., populations with a luminosity density $L_{\nu_\mu}^\mathrm{eff}$
fixed to an effective value (as a proxy for the full luminosity
function). 
In our distributional study, the average number of sources producing
$C\ge k$ track or shower events could easily be computed by convolving
the detector response (cf. \mysecref{sec:PC/formalism}) over the single-source
distributions of \mysecref{sec:P1F}; but this quantity would not fully
exploit the clustering statistics when multiple sources are present in
the same pixel, which are automatically included in the test statistic
$\mathfrak{C}$.\footnote{The number of sources contributing $k$ events
is the quantity estimated in one-point fitting studies, which use
probability-generating functions to disentangle numbers of clusters into
numbers of sources.}

\subsection{Single-pixel Likelihood Analysis \label{sec:method/likelihood}}

The quantity that we have computed to be compared to the data is the
anisotropic and spectral probability distribution of event counts.
Since there are very few events in our dataset, a $\chi^2$ analysis of
count histograms would be untrustworthy.
Here, we opt to work directly with the likelihood of the data.
The likelihood per pixel ($\mathcal{L}_p$) is a function of the number
of counts in a given pixel $p$ and in a given energy range $\Delta E_\nu$
\cite{Zechlin:2015wdz}.
Therefore, the total (binned and marginalised) likelihood of a one-point
analysis is:
\begin{equation}
\mathcal{L} = \prod_s^\mathrm{signal} ~ \prod_{\Delta E_\nu}^\mathrm{energy~bin} ~\prod_{p}^\mathrm{pixel} P(C=d^{(s,\Delta E_\nu,p)}|{\bf M})\ , \label{eq:binnedlikelihood}
\end{equation} 
under the assumption that all the count data $d^{(s,\Delta_\nu E,p)}$ in
each of the pixels, energy bins, and signal types/topologies are
mutually independent.
If detectors other than IceCube were considered in this analysis, an
additional product over independent instruments could also be considered
(cf. Appendix~\ref{sec:rant}).
In order to justify the assumption that signal topologies are indepedent, we explicitly do not consider the
``coincident'' event  (\#32) in this analysis.

The likelihood \myeqref{eq:binnedlikelihood} allows us to assess the
predictive power of a model.
Indeed, we can draw from $P(C|{\bf M})$ to generate mock data and the
exact distribution of the test statistic $TS=-2\ln(\mathcal{L})$ under
the null ${\bf M}$, from which a poorness-of-fit for the likelihood of
the real data may be computed as a $p$-value (cf. \mysecref{sec:method/p}). All the isotropic components in this study (atmospheric foregrounds, SB
and SF-AGN (SB), and 2FHL) contribute to {\bf M}.
One feature of this likelihood is that empty pixels (non-observations)
also carry information, and that this information is statistically
exploited as we will discuss in the next paragraphs.
In addition to tracks and showers, IceCube is in principle sensitive to
a number of $\nu_\tau$-specific topologies~\cite{TauNU}.
Events in these signal channels would almost certainly be of
astrophysical origin, and the nondetection of these topologies can set
strong upper limits on the astrophysical $\nu_\tau$
flux~\cite{tausearch:3yr}.
However, these unobserved topologies were not considered in our model of the
IceCube flavour response (cf. Appendix \ref{sec:flavaflav}) and do not contribute to $\mathcal{L}$.

By using $P(C|{\bf M})$ in the likelihood rather than just a Poisson at
the mean of $P(F)$, we automatically account for the skewness-induced
difference between the peak and mean values of the flux discussed in
\mysecref{sec:PF/discussion}. However, this skewness drives our prediction to lower counts, and an interesting effect occurs when both of the following occur: \begin{enumerate}
\item {\bf M} produces a count
distribution per pixel of the form
\begin{equation}
P(C) ~:~\{P(0) \approx 1-\epsilon~;~ P(1) \approx \epsilon~;~\mbox{rest}\approx 0\}\ ,
\label{eq:pathological}
\end{equation} 
\item {\bf M} is mis-specified, and produces a larger total
number of counts than in the real data.
\end{enumerate}
The effect of this convergence of features is that the real data can (counterintuitively)
give a smaller $-2\ln \mathcal{L}$ than any of the mock data generated
from {\bf M} itself.
This effect can then be used as a diagnostic for models that overpredict
the number of counts. In the context of a one-point fluctuation analysis, this could mean either
overpredicting the peak number of counts or overpredicting the amount of clustering (e.g., due
to an excess of unresolved point sources in the model).

\subsection{Digression on $p$-values \label{sec:method/p}}

We conclude our methodological overview with a few cautionary words about statistical
significances~\cite{goodman:pval}: the distributions of the test
statistics employed are non-parametric, and we find empirically that
they are asymmetric. Therefore we follow the prescription in \myrefref{GibbonsP} and report
the one-sided $p$-value along with the direction it deviates from the
bulk of the distribution of the test statistic. Although we do so at times out of convenience to the reader, we
recommend against converting these directed-$p$-values into Gaussian
$\sigma$'s. In fact ignoring the $p$-value from the other tail would
artificially inflate this significance and estimating it is error-prone~\cite{GibbonsP}. This is particularly true for the lower ``tail'' of our clustering
statistic $\mathfrak{C}$ (cf. Fig.~\ref{fig:clustering}), for which no such $p$-value exists.

After generating mock datasets for each model, we partition the mock
datasets into five subsets to generate the sample mean and standard
errors of the $p$ statistic subject to the limited number. Any $p$-values that are too small to
resolve within reasonable computational time are then quoted as upper limits. Since the uncertainties on $p$ are due to finite number of mock datasets
rather than to interesting physics, they are not systematically reported.

In order to extract as much information as possible, our analyses of the HESE data will typically focus on subsets of these data. The global signficance of independent $p$-values can be computed by correcting for the look-elsewhere effect with trial factors. Alternatively, these $p$-values may be combined with meta-analysis techniques. Because the difference between, e.g., $2.7\sigma$ and $2.8\sigma$ is somewhat irrelevant to our mock-data-limited and systematics-prone discussions (see above and \mysecref{sec:caveats}), we consider it is sufficient to correct our $p$-values with Bonferroni trial
factors (i.e., we multiply significances by the number of trials to estimate the post-trial significance), and we combine $p$-values testing the same hypothesis on different data subsets with Fisher's method $-2\sum_i^k \ln(p_i) \sim \chi^2_{2k}$. Since the test statistics in this study are built from the marginal likelihood $P(C|{\bf M})$, they have no explicit dependences on unknown
parameters and there is essentially no distinction between Fisherian (`classical') and Bayesian (`predictive') $p$-values.

\section{Analysis (II): Results \label{sec:results}}

In this section, we apply the statistical tools introduced in the previous
sections. We discuss upper limits on the resolvability of blazars and SFGs in IceCube, and
we perform various model-based one-point probability distribution analyses.

We have seen in \mysecref{sec:PC/discussion} that our astrophysical+atmospheric models produce $\mathcal{O}(20)$ neutrinos less than the HESE data.
Given the limited neutrino dataset to which we're comparing the model, and all the caveats to be specified in
\mysecref{sec:caveats}, we invite the reader to think of the following
exploratory analyses first and foremost as proofs of
concept for the methods.

\subsection{Detectability of star-forming galaxies and blazars as point sources \label{sec:results/rslvptsrc}}

The diffuse backgrounds of unresolved sources are an intrinsic and
inescapable feature of any abundant astrophysical population observed
with low angular resolution, but population self-backgrounds are not the
end of the story.
Indeed, sources visible over the \emph{total} astrophysical diffuse flux
need to be far brighter and, because flux distribution tails are
power-law-like, such bright sources are typically rather rare.

Since the angular resolution is the determining factor in this self-background effect, the number of sources we expect to resolve depends on this angular resolution. Our discussion of each source class' resolvability prospects needs to address tracks and showers separately. In this forecast we focus on spectral intensities at $E_\nu=100\;\mathrm{TeV}$, in order to evade the atmospheric backgrounds more prominent at lower energies, while maintaining a reasonable SFG contribution (spectrum of $\Gamma=2.2$) relative to the blazar contribution (spectrum of $1-s\approx0.65$). The diffuse contributions of the components are summarized in Table~\ref{tab:truncMLE}.

\begin{table}[h]
\caption{Parameters of the diffuse astrophysical neutrino flux peaks at 100 TeV, in showers and in tracks and in
 units of $10^{-20}~\intensityunit$. Note that the mean
 contribution $\hat{\mu}$ in each population is slightly larger for showers than tracks,
 while the standard deviation $\hat{\sigma}$ in tracks is wider than
 showers, as discussed in \mysecref{sec:PFC}. The $3\sigma$ and $5\sigma$
 self-background exceedance probabilities per pixel for each subpopulation
 are also reported.}
\centering
\begin{tabular}{|c|cccc|}
\hline
100 TeV & & Showers & ($\theta \sim 30\degree, N_\mathrm{pix}=48$) & \\
\hline
Population & $\hat{\mu}$ & $\hat{\sigma}$ & $>\,3\sigma$ & $>\,5\sigma$ \\
\hline
2FHL & 2.60 & 0.206 & 43\% & 32\% \\
SF-AGN (SB) & 10.61 & 0.024 & 10\% & 1.5\%\\
SB & 6.40 & 0.016 & 6.7\% & 1.1\% \\
\hline
(All) & 19.6 & 0.25 & \nonumber & \nonumber \\
\hline
\hline
100 TeV & &Tracks & ($\theta \sim 1\degree, N_\mathrm{pix} \sim 5\times 10^4$) & \\
\hline
Population & $\hat{\mu}$ & $\hat{\sigma}$ & $>\,3\sigma$ & $>\,5\sigma$ \\
\hline
SF-AGN (SB) & 10.33 & 0.84 & 2.5\% & 0.25\% \\
SB & 6.15 & 0.48 & 4\% & 0.4\%\\
\hline
(All) &16.48 & 0.97 & \nonumber & \nonumber  \\
\hline
\end{tabular}

\label{tab:truncMLE}

\end{table}

\subsubsection{Star-forming galaxies \label{sec:results/rslvptsrc/SFG}}

The expected number of SFG localised excesses resolvable above the diffuse background at 100~TeV, by a detector
with an infinite exposure and with the IceCube angular resolution for tracks are
illustrated in Fig~\ref{fig:Nsigma}. In the real data these point sources must also be extracted from the background of other (unmodeled) extragalactic contributions, and the
atmospheric foregrounds, which in our model shine an order of magnitude
brighter than all SFGs combined at 100~TeV. The number of such localised excesses is $N_\mathrm{pt} =17.8\pm4.2$ for SF-AGN (SB) and
$N_\mathrm{pt} =4.9\pm2.2$ for SB for a $3\sigma$ threshold. The fact that we forecast resolving more SF-AGN (SB) than SB is related to the locations of the $P(F)$ distributions (cf. \mysecref{sec:PF/discussion}), or more precisely, the locations of the tails of these distributions. In Fig. \ref{fig:variousPF} we see that, above intensities of about $-18.75~\mathrm{dex}~\intensityunit$, the SF-AGN (SB) tail dominates over the SB tail, and so if there are any such bright sources at all, they are more likely to be from the SF-SGN (SB) subpopulation.

\begin{figure}
\centering
\includegraphics[scale=0.75]{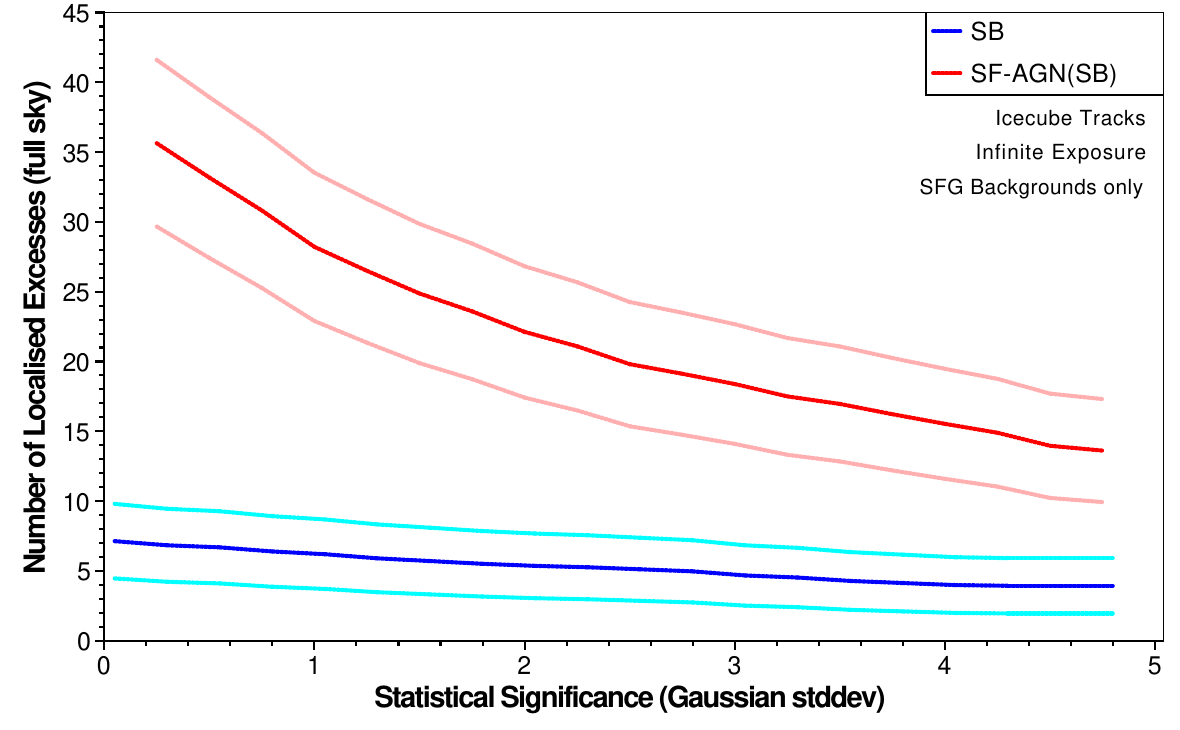}
\includegraphics[scale=0.75]{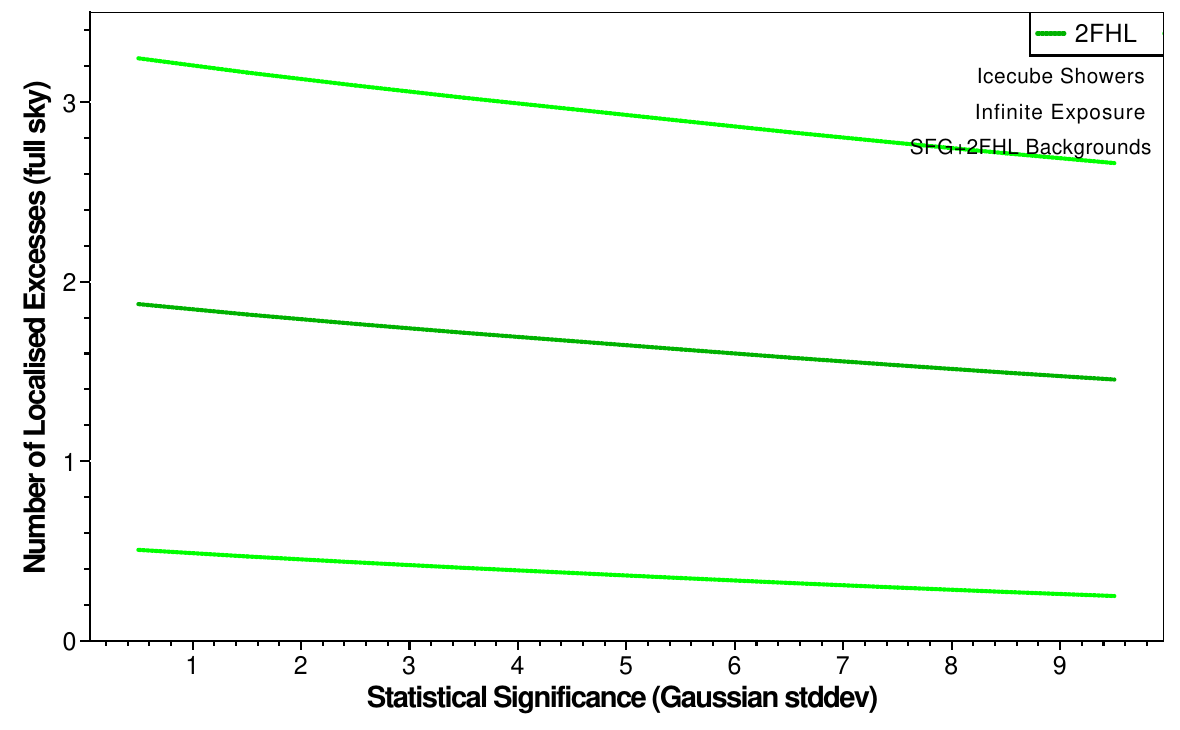}
\caption{{\it Top}: Point-source detection prospects ($N_\mathrm{pt}$ vs $\mathrm{SNR}$) for SB (blue) and SF-AGN (SB) (red) in 100 TeV tracks assuming 
 infinite exposure, IceCube angular resolution, and no backgrounds other
 than the self-background from SFGs themselves. Poisson ($1\sigma$) error bands on $N_\mathrm{pt}$ are given. These detection prospects are thus intrinsic
 and conservative upper limits. {\it Bottom:} Detection prospect upper limit for 2FHL in 100 TeV showers, assuming backgrounds from 2FHL and SFG.}
\label{fig:Nsigma}
\end{figure}

In showers, the expected number of SFG excesses over the SFG+2FHL background is essentially negligible given the order-of-magnitude difference between the estimated standard deviations of the diffuse backgrounds in Table \ref{tab:truncMLE}. A Monte-Carlo estimate suggests that we might see $N_\mathrm{pt} \sim 10^{-2}$ excesses with a negligible significance of $10^{-4} \sigma$ due to SF-AGN (SB); and no excesses due to SB, which have both a smaller $\hat{\mu}$ and a smaller $\hat{\sigma}$.
We conclude that SFGs are an intrinsically diffuse background with $30\degree$ pixels, even with an infinite
exposure.

These non-detectability claims are energy and model dependent, but finite detector
exposures and discrete neutrino events would further deteriorate the
point-source detection prospects. The number of plausible associations with SFGs~\cite{Emig:2015dma,fangkoteramurase} is bounded from above: we should not expect any corroboration of claimed associations with future data. The SFG non-detectability in IceCube should be expected also from
similar studies in gamma-rays: A one-point-fluctuation study of
Monte-Carlo simulations of unresolved blazars and SFGs in \emph{Fermi}
(which has an angular resolution comparable to that of IceCube tracks)
finds that blazars are fitted by a diffuse unresolved point source
template, while SFGs are absorbed into a diffuse isotropic template
\cite{Lisanti:2016jub}.

In short, our model makes two predictions for SFGs due to self-backgrounds
effects: firstly, SFGs constitute a diffuse background in showers;
secondly, the detectability for SFGs in tracks is still very poor. We might
see $N_\mathrm{pt}\sim\mathcal{O}(25)$ out of the $N_\mathrm{tot} \sim
\mathcal{O}(10^8)$ sources predicted from {\it{Herschel}} SB and SF-AGN
(SB) luminosity functions, and this prediction needs to be further
tempered by unaccounted-for extraterrestrial and atmospheric backgrounds
and the finite IceCube exposure, especially at energies different from 100 TeV.

In the light of these results, we draw attention to the SFG
cross-correlation programme pursued in the literature
\cite{Anchordoqui:2014,Moharana:2016mkl,Emig:2015dma}.
We have quantitatively shown that SFGs are most probably incapable of
acting as localised excesses in IceCube, even if access to far more data
than currently available were possible.
A cross-correlation of IceCube data with SFG catalogues, which relies on
such excesses, is essentially guaranteed to produce a null result
(except when a significant correlation is spuriously driven by
fluctuations or non-SFG contaminations).
This is consistent with the
null~\cite{AhlersHalzen,Anchordoqui:2014,Aartsen:2016oji} or
statistically insignificant ($p\sim0.3$--0.5
post-trials~\cite{Moharana:2016mkl}) results obtained when attempting to
correlate these high-energy events with SFGs.

Since our prediction of a null result is a function of the number of
SFGs per pixel, the only way around such negative predictions is to wait
for large quantities of data from a neutrino telescope with an angular
resolution significantly better than the $\sim$1$\degree$ achieved in
IceCube tracks.\footnote{Note that the point-source detection prospects forecasted in \cite{fangkoteramurase}, in which psf-smoothed samples of $P_1(C) \sim \delta(C- L_\mathrm{eff}(z) n_s \times \mathrm{constant~exposure})$ for a single population of sources describing all cosmic ray accelerators were used to approximate samples of $P(C)$, illustrate (qualitatively) how self-backgrounds decrease with the angular resolution also in detectors with finite exposure.} Sub-degree angular resolutions for tracks, as expected, e.g., for IceCube-Gen2~\cite{ICGen2} and KM3NeT
(ARCA)~\cite{KM3NeT}, may allow the nearest SFG point sources to be detected~\cite{fangkoteramurase}.
However, the Galactic foregrounds for such a detection in ARCA will be
significant. Note also that stacking the pixels of prospective SFG sources (as
discussed in \myrefref{MuraseWaxman}) increases the effective pixel
size, exacerbating this self-background effect (as discussed for blazars
above).

\subsubsection{Blazars \label{sec:results/rslvptsrc/2FHL}}

As discussed in \mysecref{sec:PFC}, essentially none of the 2FHL
sources contribute to their own diffuse background in tracks: all of the modeled 2FHL sources are resolvable as localised excesses in
our infinite-exposure, high-resolution detector.
This does not, however, mean that they can all be resolved as individual
objects given the backgrounds and shot noise in IceCube.
Also, this does not even guarantee a statistical detection of these sources.
Indeed, when stacking the $1\degree\times 1\degree$ muon tracks of $\sim
900$ potential blazar sources in the 2LAC catalogue~\cite{Glusenkamp},
the effective pixel size is similar to that of a $30\degree \times
30\degree$ shower, and so the effective $P(F)$ of the stack resembles
that of a single shower pixel, the self-background effect becoming
relevant again.

In showers, the blazar self-background effect does matter: at 100 TeV, only 32\% of the 2FHL sources can be resolved at $5\sigma$, even before accounting for other backgrounds. In our model of the astrophysical diffuse flux due to the
combination of diffuse fluxes from SB, SF-AGN (SB), and 2FHL sources, we only expect on average $N_\mathrm{pt} \sim 1.8 \pm 1.3$
excesses above the total
diffuse extragalactic background at 100~TeV (Fig.~\ref{fig:Nsigma} suggests this upper limit on $N_\mathrm{pt}$ is relatively independent on the detection significance).

Although this model does not necessarily rule out associations between
single high-energy showers and individual blazars \cite{PKS1424418,
Padovani:2014bha,Petropoulou:2015upa,Padovani:2016wwn}, it does place a strong (albeit model-dependent) upper limit
on the number of blazar associations we should expect to corroborate
by accumulating more shower data in a finite instrument such as IceCube. This upper limit could be further strengthened  by accounting for other subpopulations not considered in our model. At energies below or above 100 TeV, this upper limit in IceCube would be dominated by the atmospheric backgrounds or shot noise respectively.

\subsection{Clustering analysis \label{sec:results/clustering}}

Applying \myeqref{eq:clusteringC} to the 53 observed high energy
events,\footnote{The coincident event (\#32) has no directional
information and was not used in this analysis.} we find $-\ln
\mathfrak{C} =  47.4$ over the full sky.
Applying $\mathfrak{C}$ to mock data generated from $P(C|{\bf M})$ then
generates the distribution of this test statistic under the null
hypothesis, shown in Fig.~\ref{fig:clustering}.
We easily see that the model produces far less neutrino clustering than
observed (typically $-\ln \mathfrak{C} \lesssim 30$).

This is {\it not} a detection of significant clustering in the data, fully consistent with the null results in anisotropy searches~\cite{Aartsen:2014ivk,Neronov:2015,Leuermann:2016oxu}. This is due to the fact that our model underpredicts the data by about 20 counts
(cf. \mysecref{sec:PC/discussion}), and with less counts per pixel
overall one should also expect less random clustering of these counts to
occur. Although we are presumably recovering a discrepancy we already knew about, notice that we are indeed exploiting the clustering properties of the Poisson shot noise of isotropic components to see it.

\begin{figure}[h]
\centering
\includegraphics[scale=0.7]{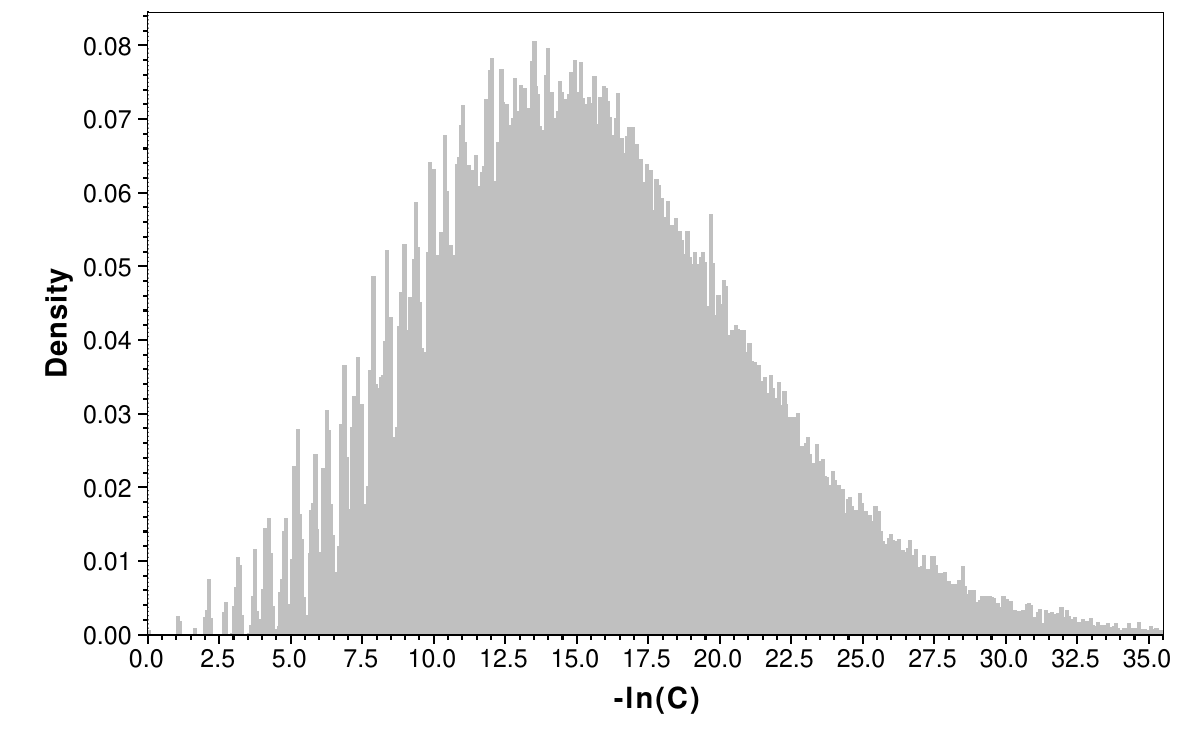}
\caption{Null distribution for $-\ln \mathfrak{C}$ applied to full-sky
 mock datasets drawn from our model (including atmospheric foregrounds,
 unresolved SFG and 2FHL point sources, and anisotropic energy-dependent
 exposure). The regular, discrete peaks (most prominent at ``low
 clustering'') are due to the finite combinatorics behind producing
 small amounts of clustering in a finite number of $30\degree$
 pixels. The value observed in the IceCube data is $-\ln \mathfrak{C} =
 47.4$.}
\label{fig:clustering}
\end{figure}

Since $\mathfrak{C} = \prod_p \mathfrak{C}^{(p)}$ is separable, this
clustering analysis can be performed on small patches of the sky. We
study the southern and northern hemispheres ($\delta \lesssim -
20\degree$, and symmetrically in the north), plus an equatorial band
($-20\degree \lesssim \delta \lesssim 20\degree$), where IceCube
effective area is maximised, to study whether the observed clustering is
consistent with that predicted by our model.
We then have $-\ln \mathfrak{C}^{(N,E,S)} = (13.7, 2.0, 31.6)$, with three trials to account for in our look-elsewhere corrections.
In the north, we find that the typical $-\ln \mathfrak{C}$ is smaller in
the mock data than in the real data, i.e., less clustering in the mock
than in the real data, but with a small one-sided $p=0.12$ ($\sim 0.7
\sigma$ pretrials).
At the equator, we find more clustering in the mock data than in the
real ones, but again with a negligible one-sided $p=0.24$ (pretrials).\footnote{The
limited number of distinct mock datasets we can generate with eight
pixels in the equatorial declination band discretises the support for
the distribution of $\mathfrak{C}^{(E)}$. This in turn generates the
bin-height alternation in the full-sky distribution of
Fig.~\ref{fig:clustering}, particularly prominent in the left-hand
tail.}

There is more clustering in the southern hemisphere of the real IceCube
data than our model could predict: our knowledge of the significance is
in this case limited by the number of Monte-Carlo realisations to the
upper bound $p<3\times 10^{-7}$ (one-sided, post-trials).
This is roughly equivalent to a $4.9\sigma$ lower limit on the
significance. This discrepancy between the discrepancies in the north and in the south is {\it not} an
observation of astrophysical anisotropy,  fully consistent with null results of anisotropy searches \cite{Aartsen:2014ivk,Neronov:2015,Leuermann:2016oxu}. Indeed, IceCube has a higher exposure in the northern hemisphere, so we expect a
larger number of counts there than in the south. We also  expect that with
the larger Poisson errors associated to this larger number of counts, the
north is more tolerant of model mis-specifications than the south (even though
all the contributions to this flux are isotropic). This interpretation
is consistent with the even less significant $p$-value in the equatorial
band, where the exposure is maximised and Poisson errors are largest.

The combined significance of these three discrepancies (according to Fisher's Method) is equivalent to $4.9\sigma$: our
data-driven model of the IceCube flux (containing only atmospherics,
SFGs, and blazars) is rejected for having less clustering than the HESE data, which is known to be consistent with isotropy. In order to accurately predict the data, one must either fine-tune the model to fit the data (by revising our extrapolations from the {\it Herschel} and {\it Fermi} data) or add additional components to the model. Since the model is still missing sources one would expect to contribute to the flux (e.g., cf. Refs.~\cite{Hooper:2016,Gaggero:2015,Zandanel:2014pva}), we believe it is premature to attempt the former (see also \mysecref{sec:caveats/model}). Updates to the model are left to future work.

\subsection{One-point fluctuation analysis \label{sec:results/1pt}}

From \mysecref{sec:PC/discussion} and the clustering analysis above, we know that the model does not produce enough neutrino event counts ($\sim 33$)
to explain the data ($=53$). But since the global likelihood (\myeqref{eq:binnedlikelihood}) is a
product of independent single-datum-likelihoods, we can decompose the
contributions of subsets of the data to our $-2\ln(\mathcal{L})$, to further
diagnose our model.

\subsubsection{Results}

We will study three energy bins wth edges at $[25,100,1000,5000]$ TeV (cf. \mysecref{sec:PC/formalism}), separately in the north and south hemispheres to fully exploit the anisotropy of the exposure. We signal-optimise away the data in the equatorial band which, as we have seen above, is least sensitive to model mis-specification. We will also decompose the likelihoods into the separate contributions from
tracks and showers; however there are not enough shower data in the northern hemisphere above 100 TeV to perform this analysis. Counting these subdivisions of the data shows there are ten trials to account for when computing global significances.

The track prediction is dominated by conventional atmospheric neutrinos
and veto-passing muons, and is surprisingly satisfactory given how
crudely we modelled the atmospheric neutrino contribution. In the south we obtain one-sided $p$-values greater than $0.3$ pretrials, suggesting no discrepancy between the model and the data. In the north, the model remains mostly consistent with the data, with a $p=0.08$ deficit below 100 TeV, a $p=0.32$ excess at intermediate energies, and a $p=0.15$ deficit above 1 PeV (all pre-trials). The combined significance of these six $p$-values is $1\sigma$ according to Fisher's method. This suggests that our
un-fine-tuned model can predict the track data fairly well, though improving the atmospheric foreground models in an attempt to extract astrophysical information out of tracks is beyond the scope of this preliminary analysis.
In what follows, only the shower data are studied to extract astrophysical information, but tracks remain
useful to the extent that they corroborate that the detector modeling and atmospheric models are correct.

\begin{table}[b]
\caption{Real/mock shower-data upper $p$-values
 (pre-trials) in the northern and southern skies and in various energy
 bands. The model includes atmospheric, SFG, and 2FHL contributions. The
 discrepancy between the mock and real data has a combined
 $\sim4.8\sigma$ significance.}
\centering

\begin{tabular}{|c|cc|}
\hline
Energy (TeV) & north & south\\
\hline
25 -- 100 & $0.218\pm 0.004$ & $(7.4\pm0.7)\times 10^{-5}$ \\
100 -- 1000 & N/A & $(1.85\pm0.3)\times 10^{-4}$ \\
1000 -- 5000 & N/A & $0.146\pm0.007$ \\
\hline
\end{tabular}

\label{tab:northsouth}

\end{table}

The results of a likelihood analysis of shower-data are summarised in Table~\ref{tab:northsouth}. The direction of the discrepancies encoded by these $p$-values confirm that we are (significantly) underpredicting the counts. At low energies, as discussed in \mysecref{sec:results/clustering}, the apparent anisotropy in the $p$-values is consistent with the difference in exposure between the north and the
south, our method being most sensitive to mismodelling in the south. This discrepancy in southern showers below 100 TeV has a significance $\sim 3 \sigma$ when accounting for the 10 trials.

At higher energies, the discrepancy between the model and the data is less severe. In the 100--1000~TeV range, the discrepancy is of a similar magnitude, $2.7\sigma$ (post-trials). At the highest energies, the hard 2FHL component is the main contribution: it underpredicts the PeV data, but only with a marginal significance of $\sim 1.1\sigma$ (post-trials). Note that our SFG model does not have a spectral break at
high energies~\cite{Tamborra:2014xia}: fixing this model shortcoming would decrease the anticipated counts
from the model, and increase the significance of the discrepancy.  At high energies in the north, there are not enough events to perform the analysis.

\subsubsection{Discussion}

The one-point analysis can also be used to ``characterise'' the discrepancy (to a first approximation). Assuming that the analysis of tracks above has validated the detector
and the atmospheric models, this discrepancy is deduced to be
astrophysical.
A further study of the energy-dependence of this discrepancy in the southern hemishepere (where our method is most sensitive to model mis-specifications) suggests that the unmodelled contribution is missing for 25--1000~TeV, but not above (cf. Table \ref{tab:northsouth}). It has a soft spectrum and/or a cutoff at high energies. We can even estimate the significance with which we need such an astrophysical component by combining the relevant $p$-values.

Combining the four $p$-values in Table~\ref{tab:northsouth} with Fisher's method (i.e., neglecting the six trials in tracks, which we know to be atmospherics-dominated) yields a global significance equivalent to $\sim 4.8\sigma$. This is only marginally better than the evidence that our model's expected number of showers ($\langle C \rangle \approx 19$) is underpredicting the data ($C=39$ showers) simply using a $\chi^2$ test, $(39-19)^2/19 \rightarrow 4.5\sigma$. The one-point analysis may not seem to add much over a standard model-based analysis, but there are a few subtleties worth mentioning here: \begin{enumerate}
\item Our computation of $\langle C \rangle$ automatically accounts for the skewness-induced bias discussed in \mysecref{sec:PF/discussion}. However, this is not the case in analyses based on $\langle I_\nu \rangle$, where the one-point skewness is ignored. Now notice that, e.g., $(39-20)^2/20 \rightarrow 4.1\sigma$. All other things equal, one-point methods based on $P(I_\nu)$ are therefore statistically stronger than analyses based on $\langle I_\nu \rangle$, because they intrinsically correct for this bias.
\item Even though we are fully exploiting $P(I_\nu)$ behind the Poisson
      shot noise, there is simply not much more information to exploit given the number of events in the HESE data. As more data becomes available, we will increasingly be able to probe the higher moments of $P(I_\nu)$, and the added statistical power of this methodology should become more apparent.
\item A likelihood approach allows us to study low-count subsets of the data where the $\chi^2$ would be unreliable. But even then, there are currently not enough showers above 100~TeV in the northern hemisphere to sensibly perform this analysis. This signal region is where one anticipates the conventional atmospheric background to contribute the least, so this analysis' potential sensitivity to a mismodelling of prompt atmospheric or astrophysical components is not fully represented in the $4.8\sigma$ combined significance.
\end{enumerate}
For a contrast of our model-based approach to other one-point fluctuation techniques, see Appendix~\ref{sec:rant}.

In summary, our likelihood analysis reveals a $\sim 4.8\sigma$ discrepancy between the model prediction for IceCube showers and the HESE data, that is especially pronounced below PeV energies. This discrepancy is insignificant (combined $1\sigma$) in tracks, suggesting it is of astrophysical (rather than atmospheric) origin in our model. The anisotropy of the discrepancy appears to be consistent with the statistical method's sensitivity to the anisotropy of the instrumental exposure.

\section{Analysis (III): Discussion of systematics \label{sec:caveats}}

The results presented above are subject to a number of caveats and uncertainties, which we discuss in this section. These fall into two categories: methodological caveats, which might introduce systematic effects; and astrophysical uncertainties, which translate into systematic uncertainties in our models.

\subsection{Methodological systematics \label{sec:caveats/method}}

Inadequacies in methodology are particularly vicious, since the biases
 they produce cannot be rigorously quantified using the tools that
 produce them. In this section,
 we discuss the two main blind spots in the single-pixel analyses above.

Firstly, the effects of extended sources that could
potentially affect our ``single pixel'' results~\cite{DMpaper} were not studied.
The instrumental point-spread function is also a relevant quantity to
consider~\cite{Zechlin:2015wdz}, as it is the energy resolution or the
difference between deposited and real
energy~\cite{ICenergyreconstruction}, amongst others. Ideally, a one-point analysis would account for these
reconstruction uncertainties at the level of the detector model, however this is far beyond the scope of this first analysis. All of these
potential systematic effects are related to our binning of the data
into energy bins $\Delta E$ and pixels $\Delta \Omega$, and need to be
addressed by (ongoing) efforts to unbin the one-pixel
functions we have been discussing into true one-point functions. This unbinning would also avoid pixelising the data with Healpix (cf. \mysecref{sec:pixsize}) when performing our clustering and likelihood analyses, freeing these analyses from pixelisation artefacts.\footnote{To check that this did not influence our results, we resampled the HESE showers within their angular uncertainties 1000 times and recomputed the clustering test statistic $\mathfrak{C}$ of \mysecref{sec:method/clustering} and the log-likelihood $-2 \ln \mathcal{L}$ of \mysecref{sec:method/likelihood}. These fluctuations do not significantly weaken these results.}

Secondly, although pixel exposure is treated anisotropically, the
incident flux distribution was assumed isotropic.
This approximation may be sufficient for studies of unresolved
extragalactic sources, but morphological, spectral, and distributional
templates will be necessary in the future to consistently account for
atmospheric and Galactic contributions.
Even for unresolved extragalactic sources, the assumption of isotropy
may be too strong, as these sources are only \emph{statistically}
isotropic. The statistical clustering of unresolved sources is indeed known to
affect the flux distributions, and in this study failing to account for
this effect underestimates the non-Gaussianity of the flux
distribution~\cite{barcons1992confusion,Barcons15061994}.

\subsection{Marginalisation systematics \label{sec:caveats/marg}}

In addition to the methodological systematics discussed above, we rely (for simplicity's sake) on the best-fit values of a number of uncertain
parameters. This
results in a likelihood that is partially profiled and partially
marginalised, and this may introduce systematics. The flux models we have adopted for the SFG and the blazars depend on data-driven parameters that remain somewhat uncertain, and using only their best-fit values is clearly dangerous when extrapolating power laws. For example, one might naively expect the $\sim$$20\%$ systematic uncertainty in the conversion $L_\gamma(L_\mathrm{IR})$ (cf. \mysecref{sec:P1F/SFG/F}) to shift the entire SFG distribution $P(F_\nu)$ in Fig.~\ref{fig:variousPF}, but if this uncertainty were marginalised away the distribution would also broaden while it shifts. A similar line of reasoning holds for the uncertainties of the luminosity function itself. There is a $\sim$$15\%$ uncertainty on the normalisation of the infrared LF for the SB subpopulation \cite{Herschel2013}, which does not affect the single-source $P_1(F_\nu)$ but does affect the multi-source $P(C)$. Even then, such $\sim$$20\%$ effects on $\approx$2.2 events from SFG in IceCube (cf. \mysecref{sec:PC/discussion}) cannot close the $\sim$20 event gap between the SFG model (driven by {\it Herschel} and {\it Fermi} data) and the HESE events, which would be inconsistent with upper limits on the SFG contribution anyway \cite{BechtolSFG}.

Another very relevant example of the mixture between statistical and systematic uncertainties in one-point methods is that the gamma-ray fluxes (and their distributions) were
extrapolated to high energies using a single value of the spectral slope
$\Gamma$ per population, rather than extrapolated with a marginalisation
over the intrinsic scatter in $\Gamma$ observed in each population.
The uncertainties on the gamma-ray spectrum $\Gamma$ are expected to affect the analysis
systematically: consider the spectral flux $F_\gamma=F_{\gamma,0} 
(E_\gamma/E_{\gamma,0})^{-\Gamma}$, with $P(F_{\gamma,0}|E_{\gamma,0})$ and $P(\Gamma)$ independent and
each approximately Gaussian.
It can then be shown that $F$ is normal-log-normally
distributed~\cite{YangNLN}.
Thus, marginalisation over $\Gamma$ generates additional skewness in
$P(F_\gamma)$, which might be used in future studies as a tool for studying
unresolved source distributions that would otherwise be treated as
Gaussians (cf. \mysecref{sec:method/rslvptsrc}).
However, in this study, keeping $\Gamma$ as a fixed parameter represents
a systematic overestimate of the gamma-ray fluxes.
It is easiest when estimating this systematic effect to ignore
distributions and look only at averages.
The mean flux of $P(F_\gamma|E_\gamma)$, assuming $\Gamma \sim \mathcal{G}(\langle
\Gamma \rangle,\sigma^2_\Gamma)$, is
\begin{equation}
\langle F_\gamma \rangle = \langle F_{\gamma,0} \rangle \times \left(\frac{E_{\gamma,0}}{E_\gamma}\right)^{\langle \Gamma \rangle+ \sigma^2_\Gamma/2} ,
\end{equation} 
in terms of the mean $\langle F_{\gamma,0} \rangle$ of an arbitrary
$P(F_{\gamma,0}|E_{\gamma,0})$.
The spectrum in our unmarginalised analysis is therefore systematically
harder than the average spectrum of the flux by a term of order $\Delta
\Gamma \sim \sigma_\Gamma^2$.
As a consequence, the predicted contributions of our extragalactic components
(extrapolated from GeV to the TeV--PeV energies) may be slightly overestimated. This may be particularly relevant for our phenomenological model of 2FHL
sources, where the mixture of source populations yields an instrinsic spread $\sigma_\Gamma$ of $\Gamma$ that compounds our choice of a harder-than-anticipated average spectrum of $\langle \Gamma \rangle=2.5$ in \mysecref{sec:P1F/2FHL}. Note that the observed spectral index uncertainty of 2FHL increases with the index itself, from $\Gamma=2 \pm 0.5$ to $\Gamma=5\pm2$ (partly because of the lower statistics) \cite{2FHL}. One might then roughly estimate the intrinsic $\sigma_\Gamma^2 \gg 0.5$.
Accounting for this effect, or not choosing a harder-than-anticipated $\langle \Gamma \rangle$, would decrease the blazar neutrino flux of the model from \mysecref{sec:P1F/2FHL}. This would presumably increase the significance of the discrepancies encountered in the
fluctuation and clustering analyses, and improve (hinder) point-source detection prospects for SFGs (blazars). However, since the $\Gamma$-marginalisation would
also broaden $P(F_{\gamma/\nu})$ the net effect would come from more than just
systematic shifts to the mean flux $\langle F_{\gamma/\nu} \rangle$.    

\subsection{Astrophysical model systematics \label{sec:caveats/model}}

Our model adopts a simplified picture of the atmospheric foregrounds, and includes only two
extragalactic source families. Both of these extragalactic models rely on extrapolations subject to astrophysical uncertainties (i.e., extrapolation of the neutrino spectra from the gamma-ray ones,  the $L_\mathrm{IR}$-$L_\gamma$ correlation adopted for the modeling of star-forming galaxies, etc.), which is inherently dangerous. Furthermore, we have illustrated in \mysecref{sec:caveats} how astrophysical uncertainties that manifest themselves as systematic shifts in averages-based methods typically also affect the shape of the one-point function when marginalised away in our distributional framework. Arguably one cannot address any astrophysical systematic self-consistently and distributionally, without incorporating the uncertainty directly into the model.

In the context of this study, we should not expect the statistical intricacies of one-point analyses to matter more than simply by changing the model to address the $\sim$$5\sigma$, $\sim$$20$ event mismatch between the model and the data (cf. Secs.~\ref{sec:PC/discussion} and \ref{sec:results/1pt}). The existence of independent upper limits on the contributions of blazars and SFGs to the flux \cite{BechtolSFG, Wang:2015woa, Glusenkamp}, that our models already saturate, suggest that it is premature to discuss upon the systematics of these subdominant contributions~\cite{FermiSFG2012,Herschel2013,Tamborra:2014xia,2FHL,FermiResolvingAbove50GeV, Padovani:2015mba}. Since we only aimed at proving the viability of our method through a simple modeling of the high-energy neutrino sky, other guaranteed sources of astrophysical neutrinos that can be well characterised using multimessenger data remain absent from the model.

In order to take into account the missing components of the neutrino flux predicted from our model, one could also consider nearby sources.
While Galactic sources are a guaranteed contribution to the neutrino flux, they are not thought to be able to produce PeV
neutrinos. However, the likelihood analysis above suggests that may not be necessary, and they can certainly generate neutrinos up to energies of a few
hundred TeV (see, e.g.,~\myrefref{AhlersMuraseGalactic} for a summary of
upper limits on Galactic contributions). Amongst other contributions, a phenomenological cosmic-ray model
designed to reconcile {\it{Fermi}}, Milagro, and local cosmic-ray data,
naturally predicts at least 10--20\% of the IceCube
flux~\cite{Gaggero:2015}, of the order of our count discrepancy.
Whether the addition of this cosmic-ray contribution to the model is sufficient to
explain the data, and a more systematic study of the model sensitivity to the
various systematic uncertainties, is left to future work.

\section{Conclusions \label{sec:concl}}

In this paper, for the first time, we explore the power of one-point statistical
analyses in the context of neutrino astronomy. Such an analysis does not require point sources to be resolved in order
to study properties of their population statistically, and, in this
sense, it is intrinsically powerful when applied to
contemporary high-energy neutrino data from the IceCube telescope.

We relied on data-driven
models of only two
extragalactic components (star-forming galaxies and blazars), besides
the atmospheric neutrino flux, and compared our predictions with the
IceCube detected flux~\cite{IC:4yr}.
The extragalactic neutrino  backgrounds have been modeled by extrapolating
multi-wavelength data from {\it{Herschel}} for the star-forming galaxy
component~\cite{Herschel2013} and from the {\it{Fermi}} 2FHL source
catalogue for blazars~\cite{2FHL,FermiResolvingAbove50GeV}.
This study has yielded three main results.

Firstly, we quantified to what extent unresolved star-forming galaxies
and blazars constitute \emph{their own background} in dedicated IceCube
point source searches.
We showed that if the neutrino flux of star-forming galaxies is well
predicted from the {\it{Herschel}} data, then star-forming galaxies are
likely to remain a diffuse, isotropic and featureless background for
IceCube: only the diffuse peak of $P(F)$ can be probed.
Note that our conclusions would be even more drastic if relying on more
conservative estimates of the SFG neutrino
contribution~\cite{BechtolSFG,Ando:2015bva}.
This model-dependent claim is unequivocally demonstrated in showers,
though in tracks we only place a conservative upper limit on the number
 $N_\mathrm{SFG} \lesssim 25$ of resolvable sources at 100 TeV (a number to be
revised in future studies due to other backgrounds and
limited exposures).
Our results are in agreement with the null results of dedicated
point-source searches and cross-correlation studies
\cite{Anchordoqui:2014,Moharana:2016mkl,Emig:2015dma,AhlersHalzen,Aartsen:2016oji}.
The opposite is predicted for blazars: if the neutrino flux of this
source population is well described by the 2FHL source catalogue, then
these sources are rare enough that self-background effects are not
relevant in tracks (see also the discussion in \myrefref{PKS1424418}).
For both source populations, these model-dependent results are consistent with one-point fluctuation analyses in
gamma-rays~\cite{Lisanti:2016jub}.

Secondly, the astrophysical distributions are found to be non-Gaussian
with power-law tails.
They are highly skewed, implying that IceCube observations are biased
away from the mean. For rare sources, the most likely and mean values
are predicted to be significantly different, by relative factors between
$0.4$ (showers) and $6.7$ (tracks) in our blazar model.
This weakens any upper limits on blazars based on the expected (mean)
contributions of these populations to the isotropic flux in tracks,
potentially by half an order of magnitude.
The skewness of the star-forming galaxy distributions is much smaller,
due to their larger abundance, therefore this effect is only percent-level.

Finally, we have applied one-point fluctuation and clustering analyses to neutrino
data. Although these analyses are model-dependent, the models we have chosen are
informed by (and otherwise consistent with) multimessenger data.
We conclude in both analyses and with a high significance that this
particular model cannot explain entirety of the IceCube neutrino events. This is not surprising, since we find
(when correctly accounting for the skewness-induced bias) that blazars, star-forming galaxies and atmospheric
foregrounds---all modeled as statistically isotropic
components---contribute in total to less than two thirds of the HESE events.
The likelihood analysis suggests that the discrepancy comes from either
systematic uncertainties on the astrophysical components or new source
populations whose spectra do not likely extend beyond 100~TeV.
Given this result and the manifest power of these one-point methods, an extended study which
takes into account more astrophysical uncertainties and more
astrophysical source populations is desirable, as it will allow convergence (even
without the need of more neutrino data) towards a multi-wavelength,
data-driven, predictive model of the high-energy neutrino sky.

\begin{acknowledgments}
We thank Daniele Gaggero, Franca Hoffman, Felicia Krauss, Jakob van Santen, and Hannes
Zechlin for useful discussions.
This work was supported by the Netherlands Organisation for Scientific
Research through Vidi grant.
IT also acknowledges support from the Knud H{\o}jgaard Foundation, the
 Villum Foundation (Project No.\ 13164) and the Danish National Research
 Foundation (DNRF91).
\end{acknowledgments}

\appendix

\section{Modeling of the IceCube effective area \label{sec:energy}}

We want to compute (distributionally) the counts registered in a pixel
due to a neutrino intensity $I_\nu = F_\nu/\Omega_\mathrm{pix}$ incident on the
detector.
Given the distribution $P(I_\nu|E_\nu)$ of the energy-differential intensity
$I_\nu(E_\nu)$, and an energy-dependent effective area $A(E_\nu)$, we want to find
the distribution $P(\mu_\nu)$ of the mean number of counts per pixel,
\begin{equation}
\mu_{\nu,\mathrm{per\,pix}} = \int_{E_{\nu,\mathrm{min}}}^{E_{\nu,\mathrm{max}}} I_\nu(E_\nu)~\Omega_\mathrm{pix} ~ A(E_\nu)~ t ~ dE_\nu\ .
\end{equation}
In what follows, subscripts $\nu$ are suppressed for notational intelligibility.

\subsection{Convolutive integration and neutrino fluxes}

We can write the integral above as a Riemann sum, i.e. 
\begin{equation}
\mu = \lim_{N \to \infty} \sum_{i=0}^{N-1} I(E+i\Delta E) A(E+i\Delta E) t \Omega ~ \Delta E\ ,
\end{equation} 
where $\Delta E=(E_\mathrm{max}-E_\mathrm{min})/N$. We see that $\mu$ is
a normalised sum $\mu = (X_0 + X_1 + \cdots + X_{N-1})/N$ of an infinite
number of random variables $X\sim I A \Omega t$, for which we might
expect the central limit theorem (CLT) to hold. These $X_i$ are not
identically distributed (so the ``classical'' CLT does not work) but
they are independent, so we might be able to use the Lyapunov CLT
\cite{petrov1975Sums}. However it is easy to show that this extended CLT does not apply either (the Lyapunov Condition is violated for our power-law tailed distributions $P(I|E)$), i.e. that $P(\mu)$ need not to be Gaussian. 
Heuristically, if our Riemann sum is $\mu \sim (X_0 + X_1 + \cdots +
X_{N-1})$ then the distribution of this infinite sum is the infinite
convolution 
\begin{equation}
P(\mu) \sim \lim_{N \to \infty} \underset{i=0}{\overset{{N-1}}{\bigstar}} P(X_i)\ ,
\end{equation} 
where we recall that $P(\cdots)$ denotes a probability density function. See Figure~\ref{fig:FAE} for a schematic of this convolution with $N=4$.

\begin{figure}
\centering
\includegraphics[scale=0.6]{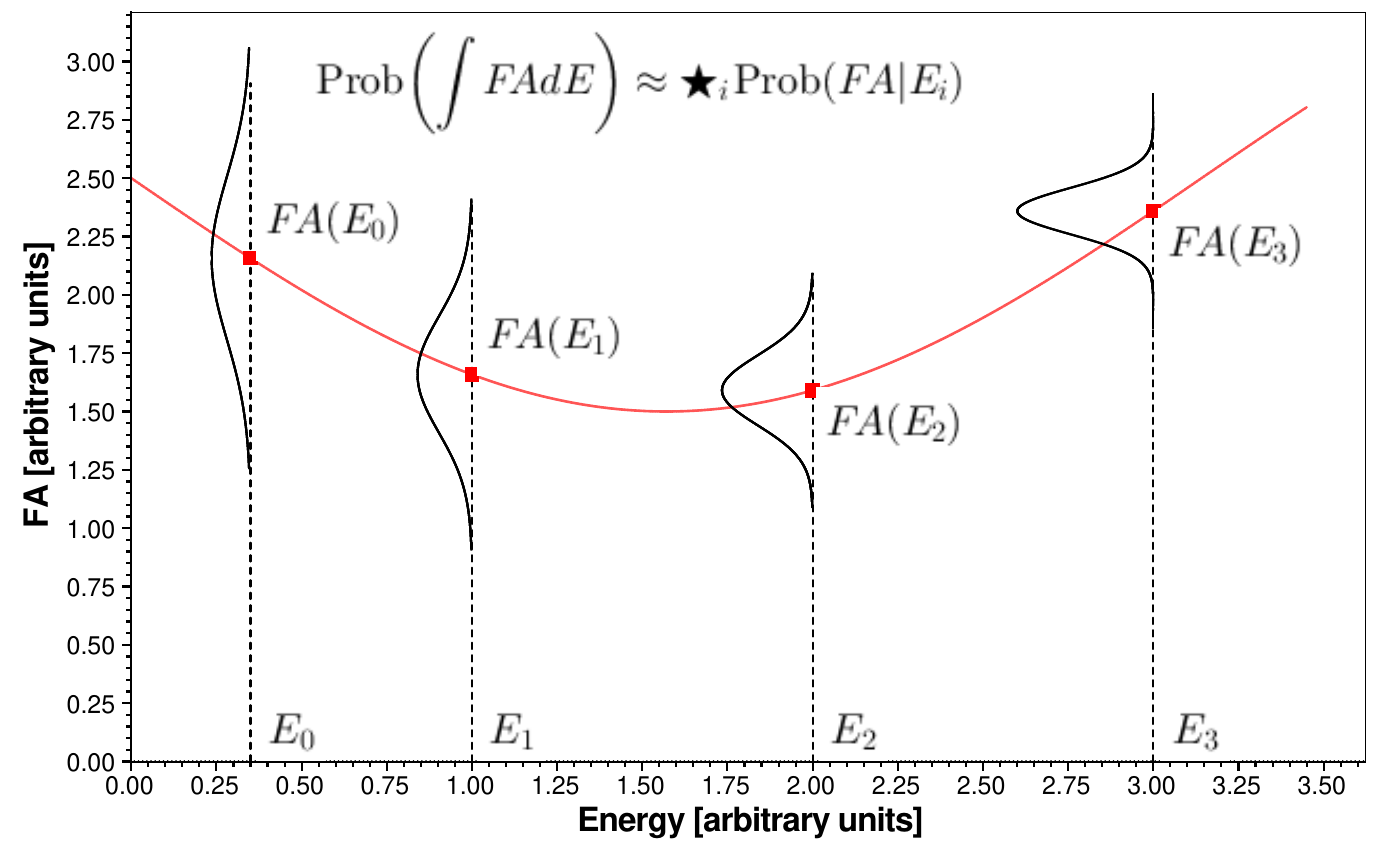}
\caption{Schematic of the integration of a conditional random variable. Specifically, this illustrates the computation of the mean-count distribution $P(\mu)$, where $\mu = \int F_\nu AT dE_\nu$ with $T$ constant and $F_\nu=I_\nu\Omega$. Since integrals look like sums, the probability distribution of an integrated quantity is the convolution of the distributions of the integrand as a function of the variable of integration.}
\label{fig:FAE}
\end{figure}

Less informally, let $\mu=\int (X|E) dE$ denote an integrated
conditional random variable (the ``primitive function'' or
``antiderivative'' of the conditional variable $X|E \in \mathbb{R}^{+}$
with respect to the random variable $E$). For our purposes, the
probability distribution function of $E$ need not be specified beyond
the fact that two fixed limits of integration $\mathcal{E}_\mathrm{min}$
and $\mathcal{E}_\mathrm{max}$ live within the support of $P(E)$. We can
then express the distribution function $P(\mu)$ as
\begin{equation}
P(\mu|\mathcal{E}_\mathrm{min}  \le E < \mathcal{E}_\mathrm{max})  = \lim_{N \to \infty}\underset{i=0}{\overset{{N-1}}{\bigstar}} \left[\delta\left(\mu-\sum_{j=0}^{N-1} X_j\right) 
P(X_i ~ | ~ E=\mathcal{E}_\mathrm{min}+i \Delta \mathcal{E})\right]\ ,
\end{equation} 
where $\Delta \mathcal{E} = (\mathcal{E}_\mathrm{max}-\mathcal{E}_\mathrm{min})/N$ and where $\delta(\mu-\sum X)$ enforces the Riemann sum on the independent summands $X_i$. This expression of course follows from the marginalisation of \begin{equation}
P(\mu,X_0,\cdots,X_{N-1}) = P(\mu| X_0,\cdots,X_{N-1}) P(X_0)\cdots P(X_{N-1}),\end{equation}
with $P(\mu|X\mbox{'s})=\delta(\mu-\sum X)$ and $N \to \infty$. A formal definition and study of this operation (presumably in terms of the It\^o-Stratonovic stochastic integral \cite{taimethod}) is left to future work, in what follows we adopt physically motivated assumptions in order to compute it. Also, it will be clearer to condense this limit of many convolutions into the notation
\begin{equation}
P(\mu|\mathcal{E}_\mathrm{min}  \le E < \mathcal{E}_\mathrm{max}) \equiv \convint_{\mathcal{E}_\mathrm{min}}^{\mathcal{E}_\mathrm{max}} P(X|E) dE\ .
\end{equation} 

Although this quantity is mathematically interesting, in practice we can not compute a number $N\to \infty$ of convolutions. Since convolution is associative, convolutive integration is composable in its boundaries:
\begin{equation}
\convint_{a}^{c} P(F|E) dE = \left(\convint_{a}^{b} P(F|E) dE\right) \bigstar \left(\convint_{b}^{c} P(F|E) dE\right)\ .
\end{equation}
Using this property and working with integrated fluxes $S=\int FdE$, we can approximate the convolutive integral of differential fluxes as the convolution of integrated fluxes:
\begin{equation}
\convint_{\Delta E} P(F|E)dE \approx \underset{i=0}{\overset{{N-1}}{\bigstar}} P(S| \delta E_i)\ .
\end{equation}
This {\it distributionally} reproduces the insight (conveyed in the main text) that the flux $S$ is an extensive quantity (with respect to $E$), so that the flux over a sum of bins is the sum of the fluxes in each bin $S(\Delta E) = \sum_i S(\delta E_i)$. With this understanding, we can finally compute 
\begin{equation}
P(\mu | \Delta E) \approx \underset{i=0}{\overset{{N-1}}{\bigstar}} P(S_i \times A_i \times t \,|\, \delta E_i)\ ,
\end{equation} where the number of convolutions is chosen large enough that $A(E)$ can be treated as a constant $A_i$ in each subbin $\delta E_i$. 

\subsection{Declination dependence \label{sec:declin}}

In addition to this energy-dependence, note that the effective area is also declination dependent. In our analysis we simply use the central declination of each pixel to compute $A(E)$. For showers, HealPix \cite{HealPix} generates pixels at seven different latitudes, calling for seven computations of $P(C)$ for each source class, for each of the two event topologies, and for each of the three energy bins. For tracks, HealPix generates 255 different latitudes. To facilitate comparisons we compute $P(C)$ at the same seven latitudes as for showers and use whichever declination is closest. This is particularly relevant for our discussion of clustering in \mysecref{sec:method/clustering}, where the distributions in tracks were coarse-grained to the scale of showers by autoconvolution.
This shortcut can only be exploited if we restrict ourselves to components with at most iso-latitudinal variations such as (to a good approximation) the atmospheric component \cite{Honda:2015}. A dedicated analysis of truly anisotropic components, such as the neutrino contribution of the Galactic plane \cite{Gaggero:2015}, is left to future work.

\subsection{Flavour dependence \label{sec:flavaflav}}

IceCube provides a separate estimation of the effective area for each of the three flavours, which we interpolate in declination and energy in order to use the formalism above. However, the effective area for tracks and showers depends on the probability $p_{T/S}^{e/\mu/\tau}$ that a neutrino of a given flavour (sampled randomly from the total neutrino flux) produces a charged or a neutral current interaction in the ice. We use the approximation \cite{PalladinoFlavaflav} 
\begin{equation}
\{ p^\mu_T=0.8,~p^\mu_S = 0.2,~p^{e/\tau}_S = 1,~p^{e/\tau}_T = 0 \}
\end{equation} 
to write 
\begin{equation}
A_{T/S} = 2 \sum_{f\in\{e,\mu,\tau\}} p^f_{T/S} \times A^f \times \eta^f\ ,
\end{equation} 
where $A^f$ is the flavour-energy-and-declination dependent quantity given by IceCube \cite{IC:2yr} and $\eta^f$ is the fraction of neutrinos of a given flavour ($\eta= 1/3$ for a $1:1:1$ flavour ratio). We multiply the effective area by 2 since the sum does not run over antineutrinos, effectively setting equal  neutrino and antineutrino fluxes. We employ a $1:1:1$ ratio for all extragalactic components, a $0:1:0$ flavour ratio for the conventional atmospheric flux, and a $1:1:0$ ratio for the prompt atmospheric flux \cite{Enberg:2008}. Percent-level atmospheric contributions from $\nu_e$ and $\nu_\tau$ fluxes (respectively) \cite{Honda:2015,Enberg:2008} are neglected, as are the neutrino-antineutrino ratios, although the fully detailed (even energy-dependent) flavour ratios can manifestly be accounted for in this type of analysis.

\section{Methodological contrast to one-point fitting \label{sec:rant}}

One-point methods, pioneered and refined by $P(D)$ analysis \cite{scheuer1957statistical,barcons1992confusion,Barcons15061994}, are currently experiencing a rebirth in contemporary astrophysics \cite{lee2009Microhalo,MalyshevHogg2011,DMpaper,Lee:2015fea, Zechlin:2015wdz, Glenn21112010, Zechlin:2016pme,Lisanti:2016jub,Vernstrom:2012,Vernstrom:2013,Breysse01102014,Breysse21032016,breyssekovetz2016probability}. The objective of such methods is typically to fit the resolved and unresolved point source distribution $dN/dF$ to the data in terms of a phenomenological model, to be interpreted after the conclusion of the analysis proper. In gamma-rays, for example, the generating function approach in \myrefref{MalyshevHogg2011} was for this purpose considerably enhanced, with careful studies of the $dN/dF$ prior-dependence \cite{Zechlin:2015wdz,Zechlin:2016pme} and systematics \cite{Lisanti:2016jub} of the fitting procedure. Both the generating-function-fitting method and the distribution-modelling method in the present study essentially require  (i) an ansatz on the source count distribution,\footnote{In the generating-function approach, this is an ansatz on the parameterisation of $dN/dF$ and on the priors associated to the parameter values.}
(ii) a model of the response of the instrument to incident flux, and (iii) a way to transition from one to the other.

The two approaches are most obviously distinguished by this third point, particularly by the direction of this transition (from model to prediction / from data to fit) and by its nature (a probabilistic hierarchical network / use of a specific statistical estimator). A perhaps more subtle distinction between the two methods is that one can model the flux distribution of many source populations in many detectors, while the specific estimator adopted in the fitting method is (at least in its current form) one-to-one. The main text illustrates the multiplicity of source populations, with different abundances, spectral indices and redshift evolutions; but besides IceCube, all instruments sensitive to high-energy neutrinos currently produce non-measurements \cite{AugerNoNu,ANITA2010,ANITA2012}. And although in this study we restrict our attention to neutrino data at the highest energies, a one-point analysis can in principle be both multi-wavelength and multi-messenger, if the model {\bf M} of astrophysics and of instrument responses that gives rise to the $P(C|{\bf M})$ count distributions is sufficiently elaborate.

In gamma-rays, there is enough data that the choice between the two methods is to a large extent a matter of taste. However, a generating-function analysis would be ill-suited to the low statistics of contemporary high-energy IceCube data: such an analysis is blind to features below the single-event sensitivity \cite{Zechlin:2015wdz}. One should expect that the experimental event count distribution is too poorly sampled to drive significant fits of the source count distribution. Even if it were not, it would \emph{by design} be incapable of disentangling the subdominant source population contributions from its unique $dN/dF$: any post-hoc interpretation of such a poorly-fit $dN/dF$ would live in a limbo of untested conjectures in wait of more data.

By contrast, in the modelling approach, various hypothetical combinations of flux distributions can be tested against the neutrino data, e.g. in terms of a likelihood ratio / Bayes factor. When they fail to be rejected by the data, or when preferences between multiple models fail to emerge significantly from the data, this occurs formally and quantifiably. On the other hand, when a model is rejected by the data, and hypotheses for this failure are put forward (in our case, the hypothesis that there is a contribution missing), these are \emph{guaranteed to be testable with contemporary data} (by improvement of the model and re-analysis). These model improvements and tests are not a methodological shortcoming, but indeed an opportunity to be explored in future work.

\bibliographystyle{JHEP}
\bibliography{Bib}

\end{document}